\documentclass{ar-1col}

\usepackage[comma]{natbib}
\usepackage{url}
\jname{Annu. Rev. Fluid Mech.}
\jyear{2022}
\jvol{55}
\doi{10.1146/annurev-fluid-120720-031006}

\usepackage{graphicx}
\usepackage[utf8]{inputenc}
\usepackage{amsmath}
\usepackage{graphicx}
\usepackage{amssymb}
\usepackage{amsmath}
\usepackage{graphics}
\usepackage{subcaption}
\usepackage{siunitx}
\usepackage{enumerate}
\usepackage{cases}
\usepackage{textcomp}
\usepackage{booktabs}
\usepackage{lineno}
\usepackage{mathtools}
\usepackage[normalem]{ulem}
\usepackage[ruled,vlined]{algorithm2e}
\usepackage{courier}
\usepackage{natbib}

\def\xb{\mathbf{x}}
\def\ub{\mathbf{u}}
\def\Ub{\mathbf{U}}
\def\cb{\mathbf{c}}

\def\ubp{\mathbf{u}^{\prime}}

\def\bhr{{\bf r}}

\def\mcG{{\mathcal{G}}}
\def\mcL{{\mathcal{L}}}
\def\mcN{{\mathcal{N}}}

\begin{document}

\markboth{Marston \& Tobias}{Anisotropic and Inhomogeneous Turbulence}

\title{Recent Developments in Theories of Inhomogeneous and Anisotropic Turbulence}

\author{J. B. Marston$^1$ and S. M. Tobias$^2$
\affil{$^1$Department of Physics and Brown Theoretical Physics Center, Brown University, Providence, RI 02912-S USA; email: marston@brown.edu}
\affil{$^2$Department of Applied Mathematics, University of Leeds, Leeds LS2 9JT, UK; email: S.M.Tobias@leeds.ac.uk}}

\begin{abstract}
Understanding inhomogeneous and anisotropic fluid flows require mathematical and computational tools that are tailored to such flows and distinct from methods used to understand the canonical problem of homogeneous and isotropic turbulence.  We review some recent developments in the theory of inhomogeneous and anisotropic turbulence, placing special emphasis on several kinds of quasilinear approximations and their corresponding statistical formulations.  Aspects of quasilinear theory that have received insufficient attention in the literature are discussed, and open questions are framed.
\end{abstract}

\begin{keywords}
turbulence, inhomogeneous, anisotropic, quasi-linear, statistics
\end{keywords}
\maketitle

\tableofcontents

\section{INTRODUCTION}
\label{introduction}

Turbulence is one of the fundamental problems of fluid dynamics yet even devising a language and mathematical framework to describe turbulence remains a formidable challenge for theoreticians --- Feynman considered turbulence as ``the most important unsolved problem of classical physics'' \citep{feyn_etal_1964}. Turbulence is ubiquitous, arising in such varied fields as engineering, geophysical, astrophysical and even biological fluid mechanics. It is difficult to overstate its importance – though many have tried – and the field has attracted attention from some of the great scientists of the last 150 years \citep[see e.g. the excellent historical perspective of ][]{Zhou2021}.

It is important at this point to state that, although turbulence is ubiquitous, homogeneous isotropic turbulence is not. Indeed, it is relatively difficult to find situations where this is the correct description of the dynamics.   In many situations of interest, turbulence is characterised by an interaction with underlying agents that naturally lead to inhomogeneity and anisotropy. In the case of geophysical and astrophysical fluid dynamics, such agents include the presence of rotation, stratification and magnetic fields – all of which engender anisotropy, with preferred directions being given by the rotation, gravity and magnetic field vectors and inhomogeneities naturally arising likewise. In engineering and biofluids, the complications may arise through the presence of systematic mean flows (e.g. for flow down a pipe and other wall-bounded shear flows) or the interactions with boundaries – both of which are difficult to capture via the traditional, elegant formalism of homogeneous isotropic turbulence. Of course, in many cases at small enough scales the fluid may forget the constraints placed upon it by the interaction with the underlying constraint, be it mean flows, rotation or stratification \footnote{We do note here that the presence of magnetic fields is more profoundly felt as one moves to smaller scales, with any cascades becoming increasingly anisotropic as the scale decreases.}, but for the larger scales it appears that a more promising avenue is to develop a theory predicated on the premise that the flows will be not be homogeneous or isotropic and that significant interactions with mean flows (and potentially mean magnetic fields) are to be expected.

\begin{figure*}
\includegraphics[width = 0.9\textwidth]{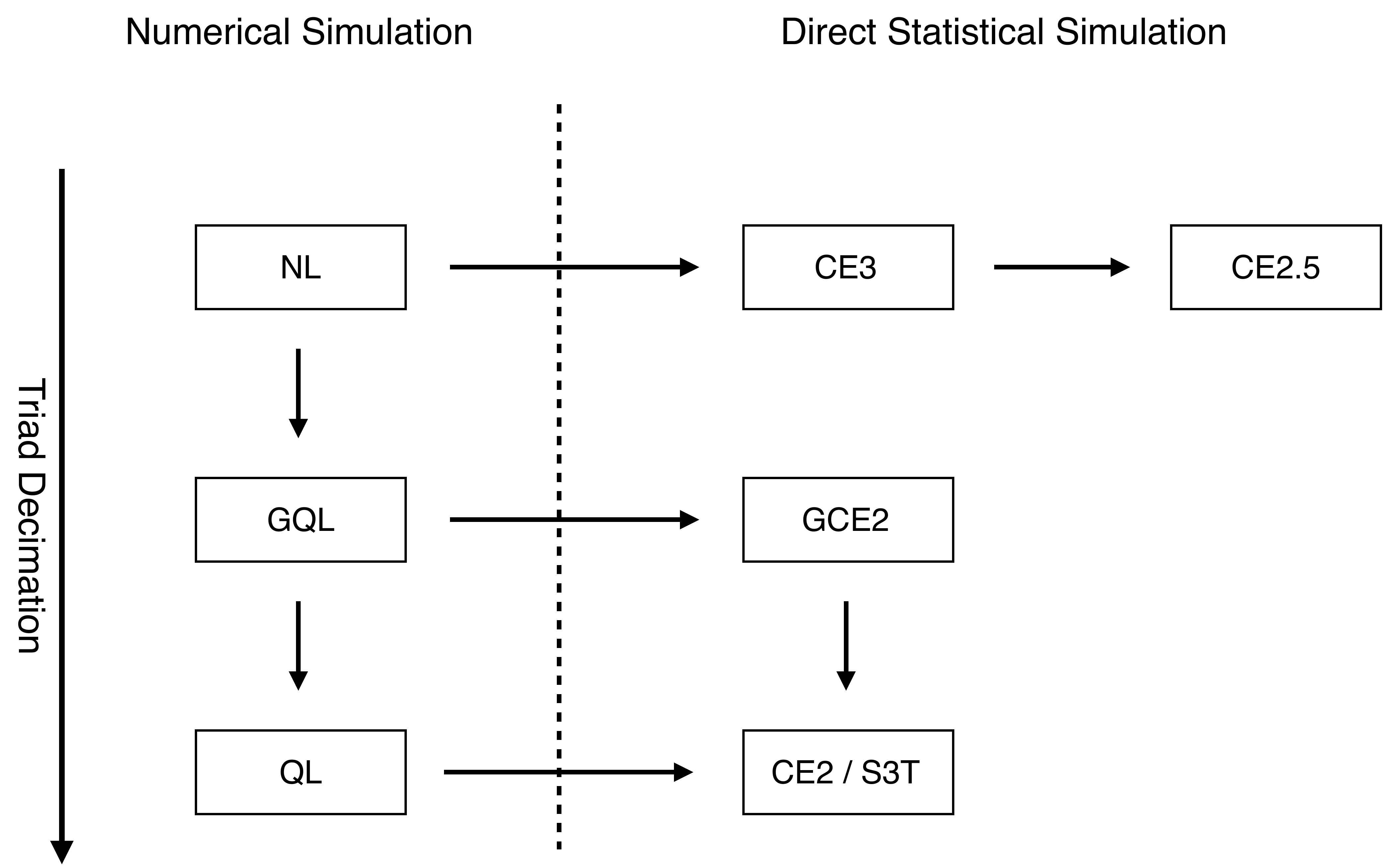}
\caption{Hierarchy of approximations descending from fully nonlinear (NL) numerical simulation that are discussed in this article. Corresponding forms of Direct Statistical Simulation (DSS) appear on the right side of the schematic. For the definition of the other terms that appear, see the text or the Terms and Definitions listed in Section \ref{terms}.  An example of each of these approximations is presented at the end of the article in \textbf{Figure \ref{fig:DSS-Comparison}}.}
\label{fig:Approximations}
\end{figure*}

For this reason, this will not be a typical and certainly not a complete review of turbulence theory and computation. For this the reader is directed to the many excellent books \citep{batchelor1953,frisch1995turbulence,pope_2000,davidson2015turbulence} or substantial reviews \citep{yag_1994,fal_etal_2001,es_2006,Zhou2021}. The article will focus instead on those developments that are concerned with a statistical description of inhomogeneous and anisotropic turbulence. We stress at the start that methods that work well as a representation of such flows – because the interaction of the turbulence with the mean flow dominates – may perform badly as a description of the idealised case so beloved by theorists; we view this as a feature rather than a bug. \textbf{Figure \ref{fig:Approximations}} gives an overview of a number of the theories and approximations that we discuss below. {Direct Statistical Simulation (DSS) -- solutions of the equations governing the statistics themselves, rather than their accumulation in numerical simulation as is usually done  -- is the focus of the later part of the article.}  Many of these methods involve the consideration of quasilinear dynamics and statistics, which is where we start our review.   

\section{QUASILINEAR DYNAMICS}
\label{QL}

We provide a high-level perspective on the quasilinear approximation and highlight some underappreciated aspects.  We also highlight some of the many open questions.  The reader is encouraged to consult the cited literature for details.  

\subsection{Historical Perspective}
\label{history}

The quasilinear approximation has its early roots in the linear Rapid Distortion Theory of Batchelor and Proudmann \citep{Batchelor1954} (RDT); for a review see  \citep{Hunt1990}.  Around the same time, Willem Malkus, who had worked as a particle physicist under Enrico Fermi before switching fields to fluid mechanics, attempted to understand how much heat could be transported by convective flows in a quasilinear like approximation \citep{Malkus:1954dh}.  The Malkus paper stimulated work on the quasilinear approximation in the early 1960s \citep{Ledoux:1961ci,Spiegel1962,Herring:1963}.  Quasilinear theory appears to have been independently discovered in the context of plasma physics \citep{Fried1960, Vedenov:1961us, Noerdlinger1963} and the phrase ``quasilinear'' may have first appeared in this context \citep{Fried1960}.  The approximation shares features with other mean-field approximations used in physics such as Hartree-Fock theory and molecular field theory and it seems possible that the advent of these quantum and statistical mean-field theories stimulated the development of quasilinear approximations. We leave this question for future historical exploration.  

Diagrammatic representations of the quasilinear approximation may be found as early as 1961 \citep{Vedenov:1961us,Vedenov:1963jj}.  Hasselman discussed the use of Feynman diagrams for wave--wave interactions in 1966 \citep{Hasselmann:1966hj}.  {A recent review of weak wave turbulence may be found in \citep{Connaughton2015}.}  Another early insight was by Herring who noted that the statistics of quasilinear equations of motion close at second-order: ``The discarding of the fluctuating self-interaction then corresponds to closing the system of moment equations by discarding the third order cumulants'' \citep{Herring:1963}. We will explore facets of Herring's observation later in Section \ref{CE2} including a surprising disagreement between the quasilinear approximation and its second-order closure.  Here we note that second-order closures began to receive renewed attention with the publication of \citep{farrellioannou2003,farrellioannou2007} for stochastically-driven barotropic jets and \citep{marstonetal2008} for a deterministically-driven point barotropic jet.  

\subsection{Choices for Averaging}
For concreteness we focus on equations of motion with quadratic nonlinearities (typically due to advection) here. The partial differential equations may be written: 
\begin{equation}
    \partial_t \ub = \mcL[\ub] + \mcN[\ub, \ub].
    \label{NL}
\end{equation}
We proceed by utilizing a standard Reynolds decomposition of the state vector into its mean and fluctuating parts; i.e.\ we set
\begin{equation}
\ub = \overline{\ub} + \ub^\prime,
\end{equation}
where the selected average satisfies the Reynolds rules of averaging so that 
$\overline{\overline{\ub}}=\overline{\ub}$,  
$\overline{\ub^{\prime}} = 0$ 
and $\overline{\overline{\ub}~ \ub} = \overline{\ub}~ \overline{\ub}$.
Many averaging procedures, such as spatial and temporal averages as well as ensemble averages satisfy these rules. Once an averaging procedure has been chosen then it is appropriate to denote mean quantities for the variables with an overbar, and fluctuations (or eddies) with primed variables.  Spatial averaging is perhaps the most common choice, with the the choice of averaging direction(s) typically dictated by the symmetries of the system. For large-scale planetary flows, averaging is generally over the zonal direction. 

\begin{figure*}
\includegraphics[width = 0.8\textwidth]{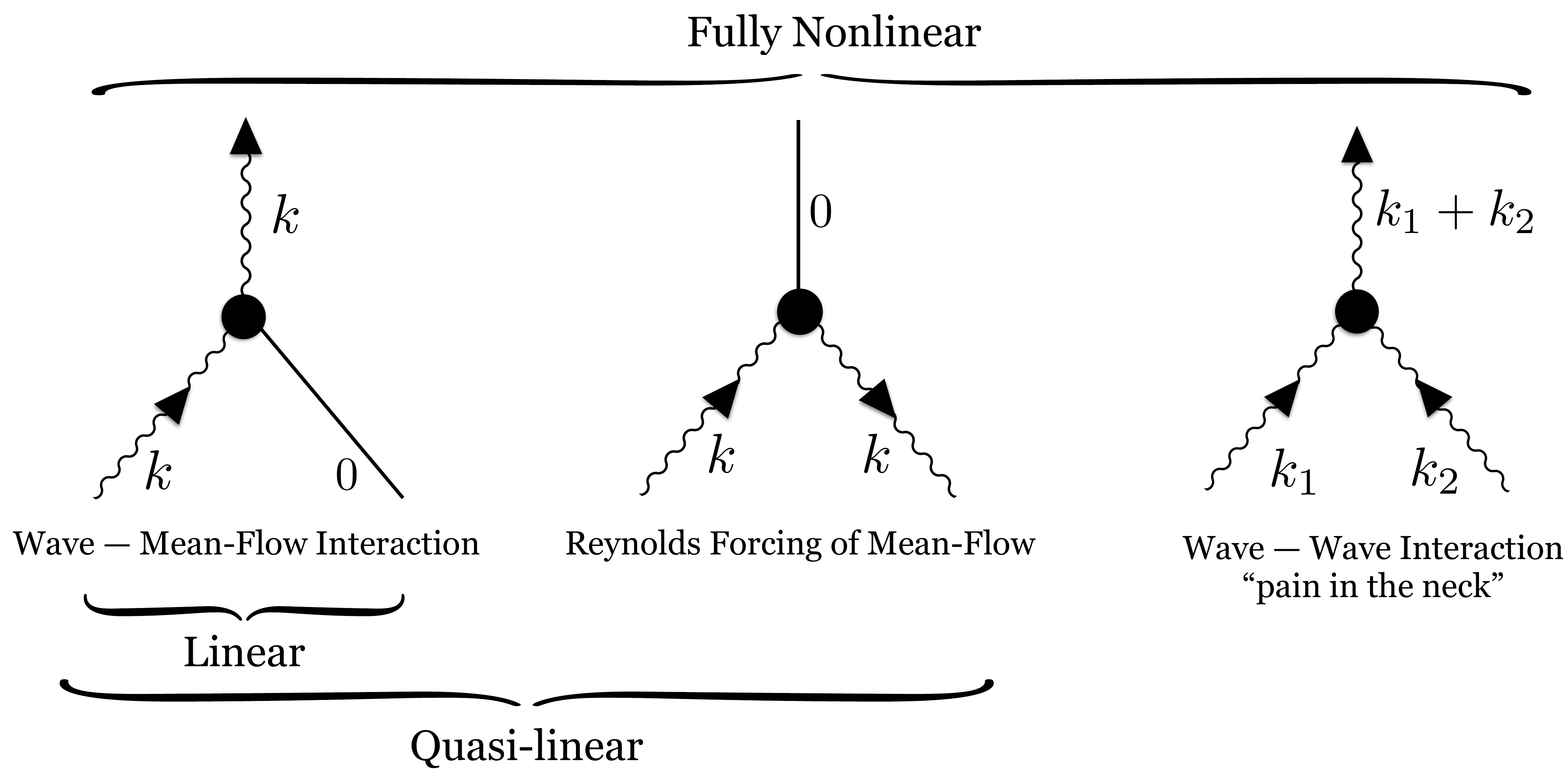}
\caption{Triad interactions in wavevector space retained in linear, quasilinear, and fully nonlinear dynamics for the illustrative case of spatial averaging over the one direction.  The solid lines represent the mean flow with wavenumber 0; curvy lines have wavenumbers that are conserved by the interaction.}
\label{fig:QL_Triads}
\end{figure*}

\textbf{Figure \ref{fig:QL_Triads}} shows the triadic interactions in wavevector space that are retained and discarded in the approximation for the illustrative case of spatial averaging for which zonal wavenumber $k = 0$ is the mean.
Ensemble averaging often makes for a better quasilinear approximation than spatial averaging because the ensemble mean-flow need not be purely zonal and hence can scatter waves into high wavenumber modes \citep{Constantinou:2016fp,Allawala:2020kq}. However ensemble quasilinear simulations are expensive, and their statistical closure suffers from the ``curse of dimensionality'' as two-point correlations are a function of both coordinates, and not their difference.  We discuss time-average means below in Section \ref{time-independent}.  

In recent years the quasilinear approximation has been explored in many different contexts.  Srinivasan and Young made a careful study of jet formation when the barotropic flow is stochastically forced \citep{srinivasanyoung2012}.  Constantinou, Farrell and Ioannou showed that such flows can reproduce jet emergence and transience \citep{Constantinou:2013fh} (see \textbf{Figure \ref{fig:MergingIntermittentJets}}).  O'Gorman and Schneider studied an idealized model of the atmospheric circulation and found that QL could reproduce some but not all of the features of fully nonlinear dynamics \citep{gormanschneider2007}. Convection in the atmospheric boundary layer was studied in \citep{ait2016cumulant} which also surveys other uses of the quasilinear approximation in the context of planetary flows.   

\subsection{Nature of the Quasilinear Approximation}
\label{natureQL}

The quasilinear approximation is a type of self-consistent mean-field theory.  The eddies or waves interact with the mean flow (first diagram in \textbf{Figure \ref{fig:QL_Triads}}) while the mean-flow is driven by Reynolds stress produced by eddies (second diagram).  {The nonlinear interaction between two waves that would produce a third wave is dropped (third diagram).} {The circumstances under which this approximation can be guaranteed to give an accurate description of the dynamics are outlined in the next section. In general, though, the approximation appears to be useful beyond the formal limit of applicability of the derivation.} The approximation is conservative:  in the absence of driving and dissipation, total energy is conserved as well as other linear and quadratic invariants such as angular momentum and enstrophy (depending on the problem).  Stability of flows typically follows from the conservation of these invariants \citep{arakawa1966}. 

Owing to the absence in the quasilinear approximation of general triadic interactions among eddies that represent eddy + eddy $\rightarrow$ eddy scattering processes --- sometimes called cascades or inverse cascades --- energy is typically confined to the lower wavenumbers, in particular waves that are marginally unstable in the presence of the mean-flow.  It is thus natural to expect the QL approximation to work well when the mean-flow is strong and waves and eddies are relatively weak.  In general this seems to be the case but a complete understanding the regimes of validity is lacking.  At least three asymptotically exact limits exist (as discussed below).  First, waves may be strongly damped due to large friction or other forms of dissipation \citep{marstonconoveretal2008,pmt2019}.  Second, there may be large time-scale separation between a slowly evolving mean-flow and rapidly evolving eddies \citep{bouchetnardinietal2013,Laurie:2014dn,Woillez:2017fr,Frishman:2017,Frishman:2018}.  And third it is possible to stimulate only a single wavenumber (for instance with stochastic forcing that is spectrally sharp in wavenumber space) \citep{Bouchet:2018er}.  In this case no other waves will be excited so long as the mean-flow strength remains below the threshold of instability.  In each of these limiting cases the quasilinear approximation is expected to hold exactly.  

\subsection{Asymptotic theories that lead to QL systems}
\label{asymptotic}

Once averaging has been performed and a Reynolds decomposition has been applied the equations may take the form. 
\begin{equation}
    \partial_t \overline{\ub} = \mcL[\overline{\ub}] + \mcN[\overline{\ub}, \overline{\ub}] + \overline{\mcN[\ubp, \ubp])},
    \label{u_bar}
\end{equation}
\begin{equation}
    \partial_t {\ubp} = \underbrace{\mcL[{\ubp}] + \mcN[\overline{\ub}, {\ubp}] + \mcN[{\ubp}, \overline{\ub}]}_\text{$\mcL_{\overline{\ub}}[\ubp]$} + \underbrace{({\mcN[\ubp, \ubp]}-\overline{\mcN[\ubp, \ubp]})}_\text{$\mcG[\ubp, \ubp]$},
    \label{u_fluc}
\end{equation}




Formally, the quasilinear approximation is applicable when the term $\mcG(\ub^\prime,\ub^\prime)$ is small compared with the quasilinear term $\mcL_{\overline{\ub}}(\ubp)$ in Equation (\ref{u_fluc}) for the fluctuating velocity ---  then this ``pain in the neck” term may be neglected. This occurs when the Kubo number, defined as 
$ Ku = \tau_c  u_{rms} / \ell_c $ is small; i.e. when the correlation time of the turbulence is short enough. The Kubo number is really an output of the system, being a property of the turbulence, and can only be determined post hoc. For this reason, it is of limited practical utility. However, there are circumstances for which a 
low Kubo number can be guaranteed a priori. These occur in certain asymptotic limits, usually ensured via a separation of timescales, though sometimes guaranteed by an ordering of the relative amplitudes of the mean flows and the fluctuations (as discussed above).

Perhaps the best-known examples of quasilinear turbulent interactions with means flows arise through the interaction of low amplitude wave turbulence with mean flows. One example of this is the quasi-biennial oscillation; here slowly varying winds in the lower equatorial stratosphere couple to gravity waves with a typical period of tens of minutes. The separation of timescales leads naturally to a quasilinear model \citep{plumb77} 

Another example is the generation of a streaming flow via the interaction of acoustic waves in a fluid with strong stable density inhomogeneities \citep[see e.g.][]{cmd2014}. Recently this asymptotic approach has been successfully extended to the case of a turbulent free shear flows in the presence of a strong stabilizing density stratification \citep{cmjrc_2022}; the separation of timescales is ensured by taking the simultaneous limits of small Froude number and large Reynolds number.

Turbulent driving of jets in the atmospheres of {gas giant planets such as Jupiter is frequently modeled by stochastically forced barotropic flows that} can also be described within a quasilinear formalism in the asymptotic limit of high zonostrophy parameter \citep[see e.g.][]{bouchetnardinietal2013}; this limit represents a separation of timescales between the evolution of the mean flow and that of the eddies (valid in the limit of weak driving and dissipation); for these strong jets and weak turbulence the emergence of a perfect staircase can be described by a quasilinear theory \citep{scottdritschel2012}. {(Note, however, that recent precise measurements by spacecraft of the gravitational fields of Jupiter and Saturn show that they have deep jets with more complicated dynamics \cite{kaspi2020}.)} Finally, in geophysical and astrophysical dynamo theory, often described by \textit{mean field electrodynamics}, the quasilinear version of the induction equation is obtained via the so-called \textit{first-order smoothing approximation} \citep{MoffattDormy:2019,krauraed:1980,Tobias:2021}, which is a reduction to a quasilinear theory. The approximation can lead to quasilinear generation of an electromotive force both in rapidly rotating systems such as Earth’s dynamo \citep{plumleyetal2018} and in accretion disks \citep{sb2015}.

{We stress again that the presence of a formal separation of timescales means that the quasilinear approximation is guaranteed to be asymptotically accurate. In other circumstances however, it is interesting to determine whether the dynamics may be well approximated by the quasilinear approximation; it may be that other considerations renders the ``pain-in-the-neck" term smaller than its quasilinear counterparts.}

\subsection{Infinite U(1) Symmetry}
\label{infiniteU1}

In the case of spatial averaging over one or more directions, the quasilinear equations of motion exhibit an infinite U(1) symmetry, with physical implications that are at present not fully understood \citep{zhang2019infinite}. {The symmetry replaces invariance under volume-preserving diffeomorphisms (equivalent to particle-relabling symmetry) of fully NL dynamics.}  The infinite U(1) symmetry can be seen by examining the triadic interactions that are retained in the quasilinear approximation (see \textbf{Figure \ref{fig:QL_Triads}}).  In both the linear, and quasilinear, approximations, the phase of each wave in spectral space can be rotated by an arbitrary amount at each zonal wavenumber $m$:  $q_m(t) \rightarrow e^{i \theta_m} q_m(t)$.  The fully nonlinear equations do not have this symmetry since each wave is coupled to every other wave.  As a consequence, there exist families of solutions of the quasilinear equations of motion with phase-shifted waves \citep{Pausch2019}.  Does the invariance of the quasilinear equations of motion under such phase rotations also correspond to an infinite family of conserved quantities?  

For purely linear waves there exist both Hamiltonian and Lagrangian formulations of the dynamics that permit the application of Noether's theorem to find the conservation laws \citep{Shepherd:1990bt}. It can be shown that pseudomomenta are conserved at each zonal wavenumber.  For barotropic flow on a rotating sphere with Coriolis parameter $f(\theta)$, for instance, the pseudomomenta are defined in terms of the vorticity $\zeta(\theta, \phi, t)$ resolved into zonal components $\zeta_m(\theta, t)$ and a mean-flow that is static in time $\overline{\zeta}(\theta)$: \begin{eqnarray}
\zeta_m(\theta, t) \equiv \int_0^{2 \pi} \zeta(\theta, \phi, t) e^{i m \phi} d\phi,
\end{eqnarray}
as
\begin{eqnarray}
{\cal M}_m = \int \frac{|\zeta_m(\theta, t)|^2}{\partial_\theta(\overline{\zeta}(\theta) + f(\theta))} \sin^2(\theta) d\theta 
\end{eqnarray}
Note that expressions for the pseudomomenta commonly found in the literature sum over all the zonal components; however each is in fact separately conserved in the linear approximation \citep{HELD:1987vj}. Pseudomomenta continue to be conserved in the quasilinear approximation when the mean flow is steady in time (Section \ref{time-independent}).  The question remains as to whether or not conserved pseudomomenta can be found in the quasilinear approximation when the mean state varies with time?  If the approximation can be formulated as a variational problem with either a Hamiltonian or Lagrangian structure, it should be possible to employ Noether's theorem to derive pseudomomenta.  If found, such quantities would have immediate practical application to geophysical and astrophysical fluid dynamics and possibly to weather forecasting.  

\subsection{Waves of Topological Origin}
\label{topological}

Rotating or magnetised fluids share basic physics with the quantum Hall effect, and the mathematics of topology plays a surprising role in the motion of the atmosphere and oceans \citep{Delplace:2017kq} and plasmas \citep{pmtz2020}.   For rotating fluids, there is a topological origin for two well-known equatorially trapped waves, the Kelvin and Yanai waves, connected to the breaking of time-reversal symmetry by planetary rotation. ( \citep{Parker:2020hf} is a pedagogical review of this physics).  Coastal Kelvin waves also have a topological origin~\citep{Venaille2021} and Kelvin's 1879 discovery of such waves \citep{Thomson:1880bv} likely marked the first time that edge modes of topological origin were uncovered although Kelvin was unaware of the topological nature. As the waves appear in regions of frequency-wavevector space that are forbidden in the bulk away from boundaries such as the equator or coastlines, the waves are protected at least partially from scattering.  This makes application of quasilinear theory attractive. The influence of background shear flow on such waves was studied in  \citep{Zhu2021}.  That work may provide an entry point to the inclusion of nonlinearities through use of the quasilinear approximation.

\subsection{Three Dimensional Quasilinear Models}
\label{3D}

In two dimensions with one inhomogeneous direction (such as a spherical surface) it is apparent that averaging should be considered over the homogeneous direction (e.g. the zonal direction), yielding a mean that is a function of the inhomogeneous direction only, whilst the fluctuations remain functions of both spatial directions. In three dimensions more choices may be available, both in the nature of the averaging and the choice of application of quasilinearity.

This choice is highlighted by the series of papers introducing, and considering the properties of so-called reduced nonlinear (RNL) models. These have been systematically examined for the paradigm problem of parallel wall-bounded shear flows \citep{thomas_etal_2014,thomas_etal_2015, farrell_etal_2016,pausch_2019}.
For this model problem, the system is inhomogeneous in the wall-normal direction and periodic in the two horizontal (streamwise and spanwise) directions. Because of the dynamics, the authors choose an averaging solely over the streamwise direction, yielding a two-dimensional time dependent mean flow. The linearisation takes place around this mean, and so the system is quasilinear in the streamwise direction but fully nonlinear in the other two directions. Once this linearisation is performed the model excites a small
number of streamwise Fourier modes and may compare well with the 
first-order turbulence statistics derived from full DNS models at moderate Reynolds numbers; this is achieved at significantly reduced computational cost \citep{bmg_2015}. However, the enforcement of quasilinearity in one direction does yield some unsatisfactory comparisons with DNS  including the scaling of streamwise wavenumber spectra with the
distance from the wall \citep{hern2022}. Some of the shortcomings of QL models may be alleviated by generalising the quasilinear approximation --- as described in the next section.

\subsection{Generalized Quasilinear Approximation}
\label{GQLsection}

The Quasilinear Approximation can be viewed as an extreme case of a series of approximations that removes triad interactions in pairs from the full dynamics \citep{kraichnan1985}. It is naturally extended to allow for the inclusion of self-consistent interactions of large-scale modes \citep{mct2016}. To explain the extension, we discuss a fluid system in a Cartesian domain with two homogeneous directions $(x,y)$ (with translation symmetry in those directions) and one inhomogeneous direction. The velocity and other variables are then decomposed into large-scale and small-scale modes, i.e.\ we set $\ub(x,y,z)= \ub_l + \ub_h$, where

\begin{equation}
\ub_l 
=\sum_{k=-\Lambda_x}^{\Lambda_x}\sum_{l=-\Lambda_y}^{\Lambda_y}{\ub}_{kl}(z) \,e^{i k^\prime x +il^\prime y} \quad \quad
\ub_h = \ub - \ub_l
\label{GQL}
\end{equation}
where $k^\prime = {2 \pi k}/{x_m}$, $l^\prime = {2 \pi l}/{y_m}$ and the $\ub_l$ and $\ub_h$ are termed the `low' and `high' wavenumber modes respectively. 
Hence, when $\Lambda_x = \Lambda_y = 0$ the low modes are the horizontally averaged (mean) modes and the high modes correspond to fluctuations about that mean (as for the QL approximation).

\begin{figure*}
\includegraphics[width = 0.8\textwidth]{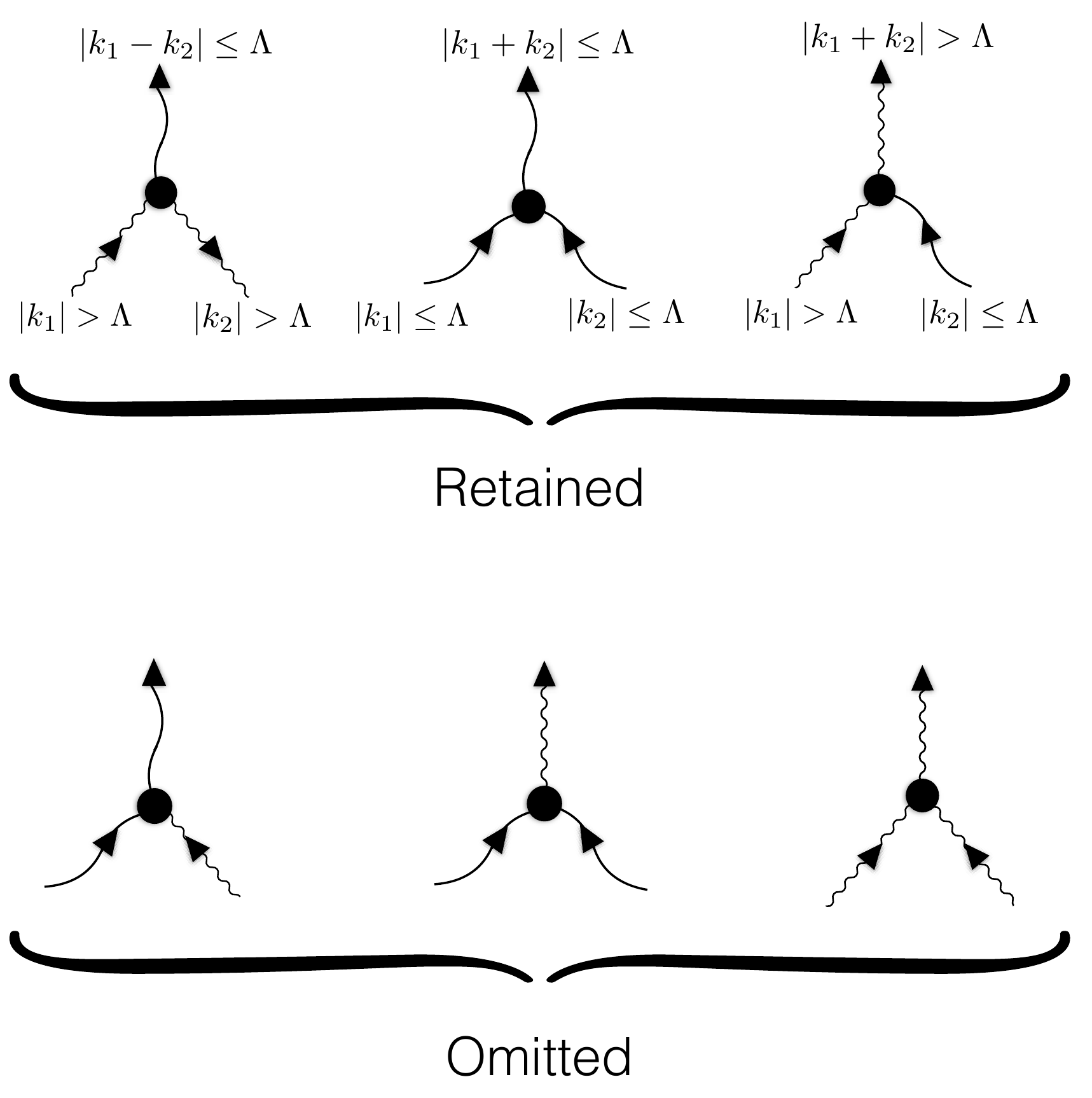}
\caption{Triad interactions in wavevector space that are retained and discarded in the generalized quasi-linear (QGL) approximation. Wavenumber cutoff $\Lambda$ separates low- and high-wavenumber modes {that are indicated in the diagrams by respectively long- and short-wavelength oscillations.  In the retained diagrams, low wavenumber modes interact fully nonlinearly, whilst high wavenumber modes only interact quasilinearly with the low wavenubmer modes. $\Lambda = 0$ corresponds to quasilinear dynamics and fully nonlinear dynamics is recovered in the $\Lambda \rightarrow \infty$.  The GQL interpolates systematically between the QL approximation and DNS.}}
\label{fig:GQL_Triads}
\end{figure*}

As discussed earlier, for the QL approximation, the removal of triad interactions in pairs is given by the interaction diagrams in \textbf{Figure \ref{fig:QL_Triads}} characterised by the suppression of certain mode interactions in the dynamics. We noted that this drastic approximation ensured that only nonlocal (in wavenumber) elements of cascades or inverse cascades were possible \citep{{tobias2011astrophysical}}. However, when $\Lambda_x, \Lambda_y \ne 0$ the Generalised Quasilinear approximation differs from QL --- the triad interactions that are retained and discarded are given in \textbf{Figure \ref{fig:GQL_Triads}}; the interactions low + low $\rightarrow$ low, high + high $\rightarrow$ low and low + high $\rightarrow$ high are retained whilst all other interactions are discarded. This selection is made so as to enable closure and preserve the relevant linear and quadratic conservation laws \citep{mct2016}. 
We note that the retention of  this set of interactions is consistent with QL when $\Lambda=0$. Furthermore as $\Lambda_x, \Lambda_y \rightarrow \infty$ the GQL system consists solely of fully interacting low modes and returns to fully NL Direct Numerical Simulation, albeit not necessarily monotonically. {In particular, if the cutoff $\Lambda$ is large (but not infinite) it is possible that high modes will be stable and (unphysically) not have any energy \citep{hern_2022a}.} 
GQL like QL is a conservative approximation but one that systematically interpolates between QL and NL.

The utility of the GQL approximation compared with that of QL has been tested on a number of paradigm turbulent fluid problems and MHD. These include the stochastic driving of jets on a spherical surface and $\beta$-plane \citep{mct2016}, three dimensional plane Poiseuille and rotating Couette flow \citep{kellam2019,hern_2022a,hern2022,tm2017}, convectively driven zonal flows in a rotating annulus \citep{tom_2018} and the helical magnetorotational instability that is crucial to angular momentum transport in disks \citep{chmt2016}. Depending on the nature of the problem --- in particular the degree of non-normality of the linear operator (see below) ---  the QL and GQL approximations may perform well or poorly in describing the statistics of the full system. However, what is clear is that in almost all cases GQL constitutes an upgrade on QL in reproducing both mean flows and low-order statistics (even for those systems that are fairly well represented by QL approximations).  The reason for this can be found in \textit{eddy-scattering}. In this process energy may be scattered between different small-scale wavenumbers (i.e. high modes) through an interaction with a large-scale flow (i.e. a low mode); for example a $k=8$ mode may scatter energy into a $k=9$ mode through interaction with a $k=1$ mode. This interaction is allowed in all forms of GQL, but forbidden in QL since there all non-zero wavenumber modes are high modes and the high+high $\rightarrow$ high interactions are removed.

Recall that for any system the fluctuations can either act to modify the (weakly dissipative) mean flow or, alternatively, interact with each other in a cascade and eventually dissipate the energy via turbulent dissipation. The partitioning of the energy between these two channels is system dependent and the ratio is linked to the Kubo number. The QL approximation assumes that all of the energy in the fluctuations is involved in interactions with the mean, and so the system saturates via a return to marginality. GQL does allow some dissipation through the non-local cascade in wavenumber space and so allows one to move away from the small Kubo number limit. The relative importance of these two mechanisms is explored via an extended Orr-Sommerfeld stability analysis in \citep{mk_2022}. 

These properties of QL and GQL lead to the hypothesis that QL dynamics can be utilised to bound transport in turbulent flows such as convection and wall-bounded turbulence. Because QL removes the eddy + eddy $\rightarrow$ eddy interactions and hence minimises the cascades that lead to dissipation, the solutions bound the transport. For example, we hypothesise that QL solutions of convection act as bounds for the Nusselt number as a function of Rayleigh number  --- numerical solutions certainly seem to suggest this. It remains to be seen how tight these bounds can be made and whether the bounding can be formalised.

\subsection{What can we learn from the case where the mean state is independent of time?}
\label{time-independent}

In this section we focus on cases where the averaging is temporal, leading to a mean state $\overline{\ub}$ that is independent of time. This is the simplest case for which to describe many of the phenomena, such as the role of noise and non-normality; though many of the considerations will carry over to other methods of averaging. Hence for the fluid equations described schematically by Equations \ref{u_bar} and \ref{u_fluc}, the mean equation simplifies to 
\begin{equation}
    0 = \mcL[\overline{\ub}] + \mcN[\overline{\ub},\overline{\ub}] + \overline{\mcN[\ubp,\ubp]}.
    \label{u_bar_steady}
\end{equation}
In this paradigm the Reynolds stress term can be thought of  as that external force 
required for the mean field to be a stationary solution.
The equation for the perturbations, which takes the form 
\begin{equation}
    \partial_t {\ubp} = \mcL_{\overline{\ub}}[\ubp] + \mcG[\ubp,\ubp],
    \label{u_fluc_schem}
\end{equation}
is again quasilinear on the neglect of $\mcG[\ubp,\ubp]$, and for a mean state that is independent of time can be solved by proposing solutions $\ubp(\xb,t) = \widehat{\ubp}(\xb) \exp (\sigma + i \omega)t$, where
\begin{equation}
    (\sigma + i \omega)\,\widehat{\ubp} = \mcL_{\overline{\ub}}[\widehat{\ubp}].
    \label{u_fluc_rzif}
\end{equation}
There exists a special class of solutions to Equation (\ref{u_fluc_rzif}) by $\sigma=0$, $\omega \ne 0$. For these solutions, the perturbations are marginally stable to the quasilinear interaction with the mean flow. Furthermore, if the basic state has the form of a periodic orbit (po) with period $T_{po}=2 \pi/\omega_{po}$, then the averaging that yields a time-independent mean is an average over a period of oscillation of the periodic orbit, and  the limit cycle is said to satisfy the RZIF property if $\omega = \omega_{po}$ (and $\sigma = 0$) \cite[see e.g.][]{Barkley_2006,turton_etal}
We stress here that equations~(\ref{u_bar_steady}) and (\ref{u_fluc_rzif}) are not closed, since the mean state has been supposed --- further elaborations to the theory are needed.

One avenue for progress is to attempt to restore some modelling of the ``pain in the neck term". This has been implemented in the resolvent analysis of \citep{mckeonsharma_2010}. In this framework the nonlinear term is regarded as input for the linear operator 
$(i \omega - \mcL_{\overline{\ub}})^{-1}$. Consider Equation (\ref{u_fluc_rzif}) with $\sigma=0$; solutions to this are in the kernel of $(i \omega - \mcL_{\overline{\ub}})$. The resolvent operator approach studies the dynamics of the singular vectors of $(i \omega - \mcL_{\overline{\ub}})^{-1}$ which are strongly amplified by this non-normal operator. Further analysis involves identifying the frequency of the  forcing that leads to the optimal growth of perturbations when supplied to the operator. In this framework the nonlinear terms in the fluctuation equation are subdominant and act solely as the source of perturbations to a singular linear operator (which is itself marginally stable to exponential growth because of the saturation of the mean flow). Fluctuations about the mean state are caused solely by transient amplification of the noise supplied by the nonlinear interactions of the fluctuations themselves. Of course the fluctuations have played a role in saturating the mean; the mean can be determined by temporally averaging a fully non-linear calculation \citep{mckeonsharma_2010} or by self-consistently solving the coupled set of equations
\begin{eqnarray}
    0 &=& \mcL[\overline{\ub}] + \mcN[\overline{\ub},\overline{\ub}] + \overline{\mcN[\ubp,\ubp]},\label{mean_again}\\
 ( i \omega -  \mcL_{\overline{\ub}})\widehat{\ubp} &=& f \exp{i \omega t}. \label{resolvent_forcing}  
\end{eqnarray}

Note the difference in philosophy of the two approaches. In the first, where the mean flow is taken as the average of a full simulation, it is known that the mean flow is, on average, marginally stable. The fluctuations calculated by optimising the growth in Equation (\ref{resolvent_forcing}) contribute to the dynamics about that mean, but do not self-consistently drive the mean. If the coupled system (\ref{mean_again}-\ref{resolvent_forcing}) is solved then the fluctuation is both optimal (in terms of transient growth) \textit{and} self-consistently drives the mean used in the operator of the resolvent analysis. However, because the nonlinear term in the fluctuation equation has been neglected the self-consistent mean flow is not that obtained from a fully nonlinear simulation of the system --- any differences can be ascribed to the replacement of the correct nonlinearity with a periodic function.

Another solution procedure that yields self-consistent fluctuations and mean flows is the SCM (self-consistent method) \citep{Mantic-Lugo:2014}. In this paradigm, one solves self-consistently for the mean flow and the leading order eigenmode of the system ($\ub_{scm}$). A return to marginality hypothesis further assumes that the resultant mean flow (${\boldsymbol U}_{scm}$) is marginally stable (and so has an eigenvalue with zero real part). The resulting system is given by
\begin{eqnarray}
    0 &=& \mcL[\Ub_{scm}] + \mcN[{\Ub}_{scm},{\Ub}_{scm}] + \mcN[\ub_{scm},\ub^*_{scm}],\label{scm_mean_again}\\
 0&=&( i \omega_{scm} -  \mcL_{\Ub_{scm}}){\ub_{scm}}, \label{scm_eqn}  
\end{eqnarray}
where $\mcL_{\Ub_{scm}}[\cdot] = \mcL[\cdot] + \mcN[\cdot\,,{\Ub}_{scm}]+ \mcN[{\Ub}_{scm},\,\cdot]$. This ansatz can be extended to higher order \citep[see e.g.][]{meliga_2017,bt_2021}. This takes the form 
\begin{eqnarray}
    0 &=& \mcL[\overline{\Ub}] + \mcN[\overline{\Ub},\overline{\Ub}] + \sum_{1\le|m|\le M}{\mcN[\ub_m,\ub_{-m}]},\label{mean_again_trunc}\\
 ( i n \omega -  \mcL_{\overline{\Ub}}){\ub_n} &=& \sum_{\substack{1\le|m|\le M \\ 1 \le |n-m|\le M}} {\mcN[\ub_m,\ub_{n-m}]}, \label{pert_trunc}  
\end{eqnarray}
for $1 \le n \le M$. Although this procedure in the temporal domain is somewhat reminiscent of the GQL approximation for spatial partitioning described above it is important to note that in this ansatz, nonlinear interactions between the perturbation terms that do not contribute to the evolution of the mean are included \citep{bt_2021}.

A final approach to modelling the nonlinear term is to consider the response of the equations linearised about the mean (in time) flow to stochastic forcing. It is now well established that such linear models with white noise stochastic excitation can yield the spatiotemporal features reminiscent of both fully-developed and transitional turbulence \citep[see e.g.][]{Farrell1993StochasticFO,Hwang2010AmplificationOC,hwang_cossu_2010}. Despite this success, it is clear that modelling the nonlinear term as a white-in-time
stochastic excitation does not reproduce the correct statistics of the fluctuating
velocity field, as perhaps might be expected \citep[see e.g.][and the references therein]{zare_etal_2017}. However, utilising the linearised equations renders the system susceptible to the advanced methods of modern robust control \citep{zare_etal_2017}, so the question of the preferred form of the stochastic driving of the linearised equations may be turned into one of optimisation, using a maximum entropy formulation
together with a regularization that serves as a proxy for rank minimization. \citep{zare_etal_2017} demonstrate that the fully nonlinear and coloured-in-time stochastically driven
linearized NS equations can be made equivalent at the level of second-order statistics, for turbulent channel
flow if a suitable coloured-in-time stochastic driving is utilised.

\begin{figure*}
\includegraphics[width = 0.8\textwidth]{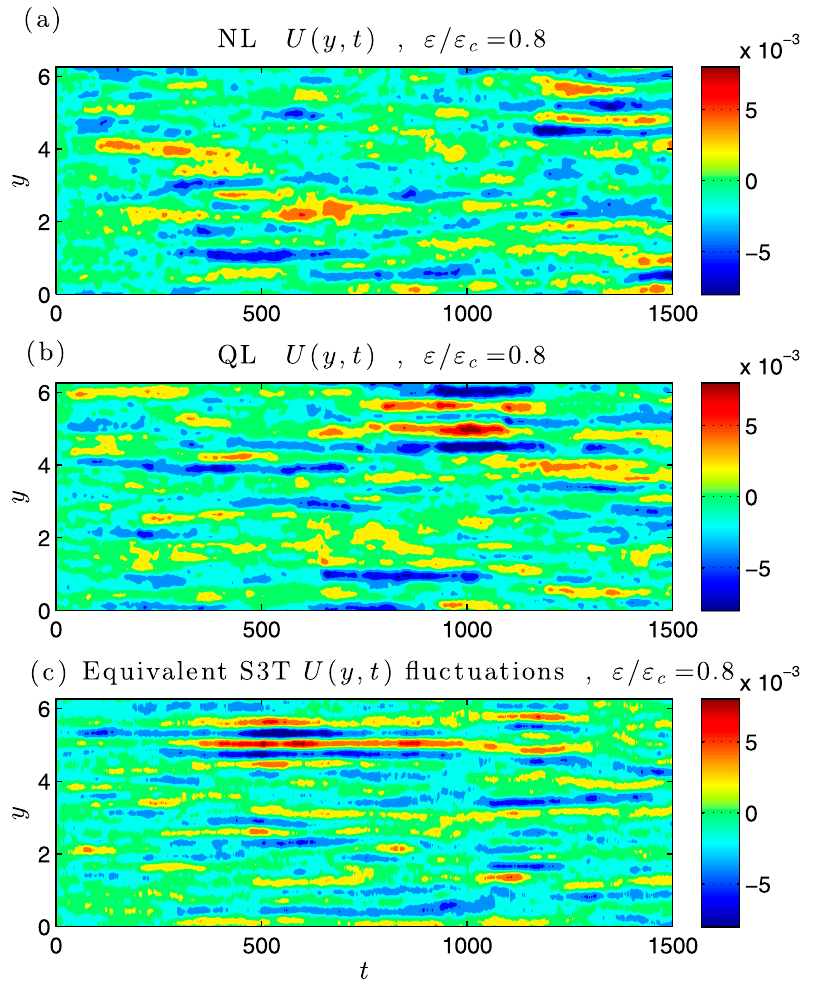}
\caption{Hovm\"oller diagrams of stochastically forced barotropic flows showing that transient jets can be reproduced in quasilinear dynamics.  Figure adapted with permission from  \citep{constantinoufarrelletal2013}   \textcopyright American Meteorological Society 2014.}
\label{fig:MergingIntermittentJets}
\end{figure*}

\section{QUASILINEAR STATISTICAL THEORIES AND DIRECT STATISTICAL SIMULATION}
\label{CE2}

Many quasilinear approximations have exact closures in terms of low-order statistical moments or cumulants, enabling their Direct Statistical Simulation (DSS) -- that is, bypassing numerical simulation of the QL EOMs to solve directly for their statistics.  {The first, second and third cumulants are centered moments.  (Fourth and higher cumulants differ from centered moments.  For example, the fourth cumulant vanishes for a normal distribution whereas the fourth centered moment does not \citep{marston2014direct}.)} Introducing the first and second equal-time cumulants
\begin{eqnarray}
    \cb(\bhr) &\equiv& \overline{\ub(\bhr)},
    \nonumber \\
    \cb(\bhr_1, \bhr_2) &\equiv& \overline{\ubp(\bhr_1) \otimes \ubp(\bhr_2)},
    \label{lowCumulants}
\end{eqnarray}
and using the identity
\begin{eqnarray}
\ubp(\bhr_1) = \int \delta(\bhr_1 - \bhr_2)~ {\bf I}~  \ubp(\bhr_2)~ d\bhr_2 ,
\end{eqnarray}
where $\bf I$ is the identity matrix, the average of the nonlinear term in Equation \ref{u_bar} may be rewritten in such a way that the two appearances of $\ubp$ on the RHS may be brought together to form the second cumulant, so that the evolution equation for the mean can be written
\begin{eqnarray}
    \partial_t \overline{\ub} = \mcL[\overline{\ub}] + \mcN[\overline{\ub}, \overline{\ub}] + \int \mcN[\overline{\ubp(\bhr_1) \otimes \ubp(\bhr_2)}, \delta(\bhr_1 - \bhr_2)~ {\bf I}] d\bhr_2 .
\end{eqnarray}
This may now be expressed in terms of the first and second cumulants as:
\begin{eqnarray}
    \partial_t \cb(\bhr_1) = \mcL[\cb(\bhr_1)] + \mcN[\cb(\bhr_1),~ \cb(\bhr_1)] + \int \mcN[\cb(\bhr_1, \bhr_2), \delta(\bhr_1 - \bhr_2) {\bf I}] d\bhr_2,
    \label{1stCumulantEOM}
\end{eqnarray}
where it is understood that $\mcN$ acts only on $\bhr_1$ and its associated vector in second cumulant Equation \ref{lowCumulants}; the second coordinate $\bhr_2$ and vector come along for the ride. 
The EOM for the second cumulant may be found by multiplying Equation~(\ref{u_fluc}) by $\ubp(\bhr_2)$ followed by averaging.  This yields:
\begin{eqnarray}
\partial_t \cb(\bhr_1, \bhr_2) &=& \mcL_{\cb(\bhr_1)}[\cb(\bhr_1, \bhr_2)] + 
\mcL_{\cb(\bhr_2)}[\cb(\bhr_1, \bhr_2)],
\nonumber \\
&=& 2 \{\mcL_{\cb(\bhr_1)}[\cb(\bhr_1, \bhr_2)]\}
\label{2ndCumulantEOM}
\end{eqnarray}
where the operator $\mcL_{\cb(\bhr_1)}$, introduced in Equation \ref{u_fluc}, acts only upon the $\bhr_1$ coordinate.  
Here we have introduced the short-hand notation $\left\{ \right\}$ for symmetrization that maintains the invariance of the statistics under interchanges of the field points $\cb(\bhr_1, \bhr_2) = \cb(\bhr_2, \bhr_1)$; explicitly, 
$\left\{ \cb(\bhr_1, \bhr_2) \right\} \equiv \frac{1}{2} [\cb(\bhr_1, \bhr_2) + \cb(\bhr_2, \bhr_1)]$. The equations of motion for the two cumulants, Equations \ref{1stCumulantEOM} and \ref{2ndCumulantEOM}, are closed because the ``pain in the neck" term has been dropped.  Had it been included, the second cumulant would have coupled to the third cumulant, the third to the fourth, and it would be turtles all the way down the hierarchy of cumulants \citep{marston2014direct}.  One might assume that closure at second order implies Gaussian statistics but this is not the case.  The decoupling of the first and second cumulants from the third and higher cumulants in the quasi-linear approximation does \emph{not} mean that the higher cumulants necessarily vanish but only that they do not affect the first two cumulants. Highly non-Gaussian statistics can appear in quasilinear approximations (see Section \ref{largeDeviation} below).  

The CE2 closure, based as it is upon quasilinear dynamics, is a conservative approximation and also realizable: The second cumulant is a positive-definite matrix \citep{kraichnan1980realizability} or more precisely positive-definiteness may be enforced. The rank instability (see Section \ref{rank}) may lead to negative eigenvalues that should be projected out to ensure stability.  
The statistical formulation has several limitations.  It is not obvious how higher-order nonlinearities such as step functions (for instance to account for latent heat release) may be incorporated.  Correlations between fields at two different times would seem to be excluded despite the fact that lagging correlations may be stronger than equal-time correlations.  Stochastic forcing may however be included in this statistical formulation by adding the covariance matrix to the RHS of Equation \ref{2ndCumulantEOM} so long as the noise is delta-correlated in time.    
 
Stochastic Structural Stability Theory (SSST / S3T), part of a program of Statistical State Dynamics (SSD), is related to CE2.  
CE2 and S3T differ, however, as S3T usually includes small-scale stochastic forcing that is delta-function correlated in time to represent the missing eddy + eddy $\rightarrow$ eddy scattering. Energy injected by the random forcing is balanced by damping. S3T is also often applied to mean flows which do not have linearly unstable modes as it focuses on non-normal growth and decay of the stable modes that are driven by stochastic forcing.  S3T has been used to study zonostrophic instabilities in barotropic flows \citep{bakas2011structural,bakas2013emergence}, the evolution of jets \citep{farrellioannou2003,farrellioannou2007,Constantinou:2013fh}, and non-zonal \citep{bakas2014theory} coherent structures.

CE2 has often been applied to flows with instabilities.  In the geophysical and astrophysical context this includes \citep{marstonconoveretal2008,marston2010,marston2012,pmt2019}.  It also takes a different approach than S3T by not attempting to parameterizing eddy--eddy interactions.  The absence of an adjustable parameter means that CE2 has greater predictive power than S3T but is also more likely to fail to reproduce NL dynamics.  These failures can be quite instructive (See \citep{tobias2013direct} and \textbf{Figure \ref{fig:DSS-Comparison}}). CE2 has been used in problems as ambitious as three-dimensional plasmas \citep{sb2015}.  The ``curse of dimensionality" becomes pressing as the spatial dimension increases from 2 to 3 and methods to tame it may be required (see Section \ref{DimensionalReduction}).  

\subsection{Methods of solution for fast/slow QL equations}
\label{FastSlow}

Having derived quasilinear equations (sometimes via an asymptotic procedure), it is sensible to exploit the structure of the equations in their solution. Usually the system takes the form of equations that admit solutions for the mean that are relatively slowly varying and equations for either the fluctuations (for QL) or for the evolution of the correlation functions (for statistical theories \citep{delsolefarrell1996}). Different strategies for solution may be needed for the two different cases, since the fluctuations evolve on a much more rapid timescale than their correlation functions (which usually evolve on the same timescale as the mean flows for a QL theory). For QL DNS, strategies may be developed that make use of the separation of timescales, either by utilizing a HMM-like solver that combines a macrosolver with a large integration timestep for the slow (mean) dynamics and a microsolver for the fast dynamics \cite[see e.g.][]{Tretiak_2022} or by exploiting the linearity of the equations for the fluctuations \citep{mc2019}. Whilst the first method is a general tool that works well for general situations where the timescales are well separated, even for the case where the fast dynamics is inherently nonlinear, the second method utilizes the linearity of the fast system, and therefore the fact that the slow field must stay close to a state of near marginality to achieve additional efficiency. For quasilinear DSS (CE2), both equations evolve on a slow manifold and large timesteps are possible using implicit Krylov subspace methods \citep{saad_2003}. Indeed, here the ultimate attractor for the statistics can sometimes be found directly using minimisation techniques \citep{li2021_l63,Li2021_dyn}, though it is unclear when these are competitive with the timestepping methods described above.

\subsection{Rank of the Second Cumulant}
\label{rank}

As discussed above in Subsection \ref{CE2} the equations of motion for equal-time cumulants close exactly at second order in the quasilinear approximation. One might be tempted to conclude that statistics thus obtained should agree exactly with those found from averaging the quasilinear dynamics. Surprisingly this turns out not to not always be the case \citep{Nivarti2022}.  Because the second cumulant has more degrees of freedom than the dynamical fields themselves (for instance in Equation \ref{lowCumulants} the second cumulant depends on two spatial coordinates) there can be instability in rank even when it is initialized to be unit rank as in Equation \ref{lowCumulants}.  For the case of zonal or spatial averaging, the rank instability may then lead to a difference in the distribution of spectral power at different zonal wavenumbers. 

\begin{figure*}
\includegraphics[width = 1.0\textwidth]{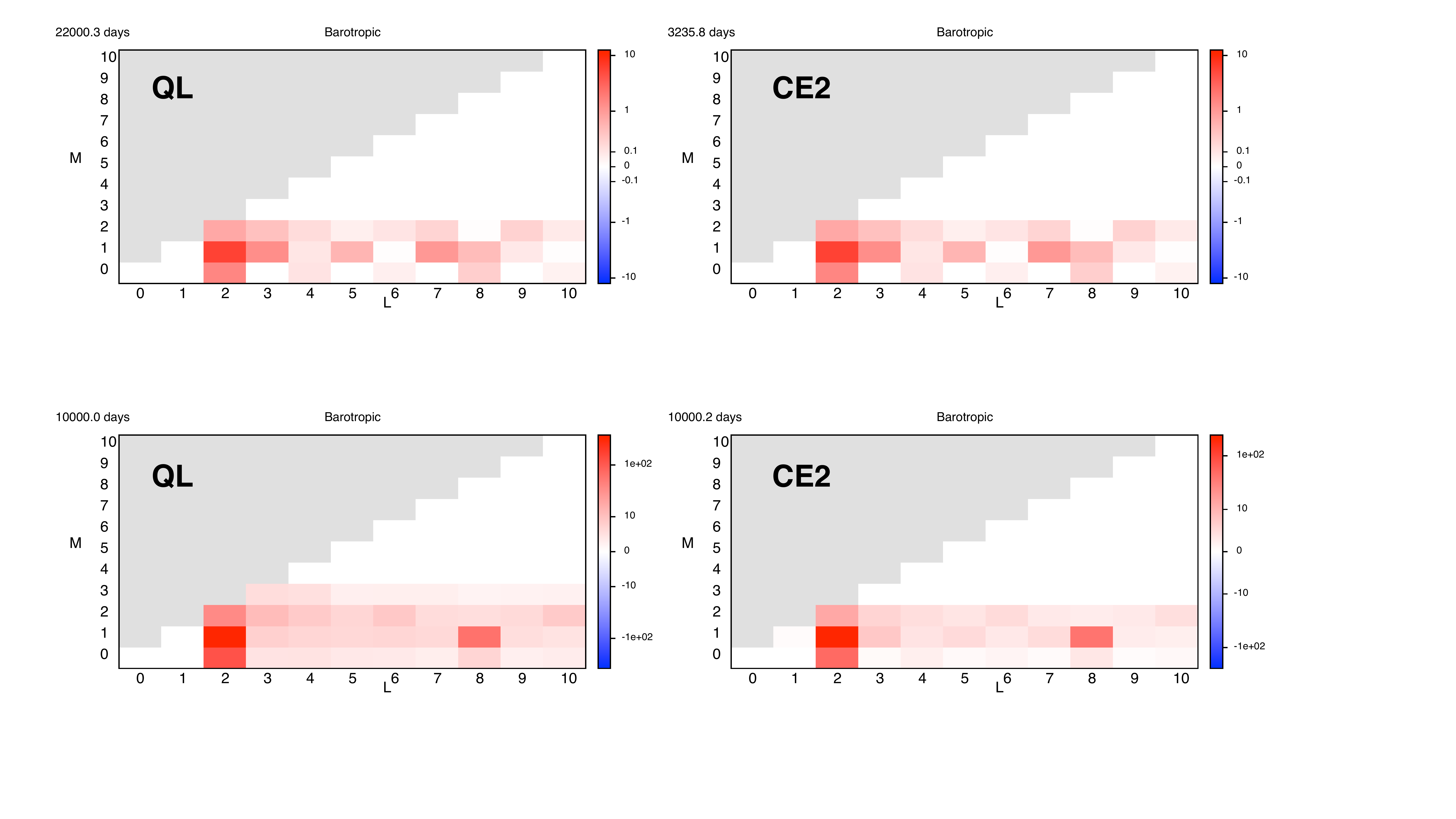}
\caption{Power spectra of QL simulations (left) and CE2 (right). The zonal wavenumber $m$ is plotted on the vertical axis, and the spherical wavenumber $\ell$ is along the horizontal axis. Top:  With weak forcing, QL and CE2 agree.  Bottom: For strong forcing, QL has energy in zonal mode $m = 3$ not found in CE2 due to a rank instability (see Section \ref{rank}).}
\label{fig:Power}
\end{figure*}

We illustrate the rank instability here on the unit radius sphere with fully spectral code\footnote{The macOS app ``GCM'' that implements the simulations shown in Figures \ref{fig:Power}, \ref{fig:Stress}, and \ref{fig:DSS-Comparison} can be downloaded at URL https://apps.apple.com/us/app/gcm/id592404494?mt=12}.  
The time evolution of the radial component of the relative vorticity $\zeta \equiv \widehat{r}\cdot (\nabla \times {\boldsymbol u})$ is given by 
\begin{equation}
    \label{eq:vorticity}
    \partial_t \zeta =  J[\psi,\zeta] + \nu \nabla^2 \zeta + F(\theta), 
\end{equation}
where ${\boldsymbol u}$ is the velocity, $J[\psi,\zeta]$ is the Jacobian on the sphere and the streamfunction $\psi \equiv \nabla^{-2}\zeta$. Kolmogorov-like forcing is chosen to be $F(\theta) = a (P_2(\cos{\theta}) + 8 P_8(\cos{\theta}))$, where $P_\ell$ are Legendre polynomials and $\theta$ is the co-latitude.  \textbf{Figure \ref{fig:Power}} shows that while there is perfect agreement between QL and CE2 for gentle forcing ($a = 0.25$), the power spectra disagree at strong forcing ($a = 1.0$) with QL showing energy at zonal wavenumber $m =3$ whereas no such energy is found in CE2.  Note that the second cumulant, as calculated from zonal averaging snapshots of the QL flow, always has a rank of $0$ or $1$ at each zonal wavenumber.  Here by contrast the rank of the second cumulant obtained from CE2 is greater than 1 for the zonal wavenumbers with energy.  

The possible presence of the rank instability for stochastically forced models, or for ensemble averaging, are questions for further investigation. Both QL and CE2 are approximations to the full dynamics so the existence of the rank instability does not necessarily mean that CE2 is less accurate than statistics obtained from QL.  Indeed the initial condition for CE2 is often chosen to have a maximal rank second cumulant (sometimes this is called the ``maximal ignorance'' initial condition) and thus it may be the case that CE2 offers an advantage over QL as the choice of initial condition amounts to a form of ensemble averaging that may be carried out alongside spatial averaging.

\subsection{Generalized Cumulant Expansion}
\label{GCE2}

An attractive feature of GQL is that, like QL, the equations for the equal-time statistics close exactly at second order.  This closure, the Generalized Cumulant Expansion (GCE2) leverages the linearity of the EOM for the high wavenumber GQL modes, given by Eqn.~(\ref{GQL}).  Therefore, in the GQL approximation, the EOM for the two-point statistic $\overline{\ub_h(\bhr_1)~ \ub_h(\bhr_2)}$ closes and can be evolved in time with no further approximation beyond the one already made in GQL.  (The low-wavenumber modes $\ub_l(\bhr)$ continue to evolve fully nonlinearly as in GQL.) GCE2 is automatically realizable because it is an exact closure of the GQL dynamics, and should give a more accurate reprentation of the true statistics than CE2. An as-yet unanswered question is to what extent the rank instability discussed in Section \ref{rank} affects GCE2.  

\subsection{Large Deviations}
\label{largeDeviation}

\begin{figure*}
\includegraphics[width = 1.0\textwidth]{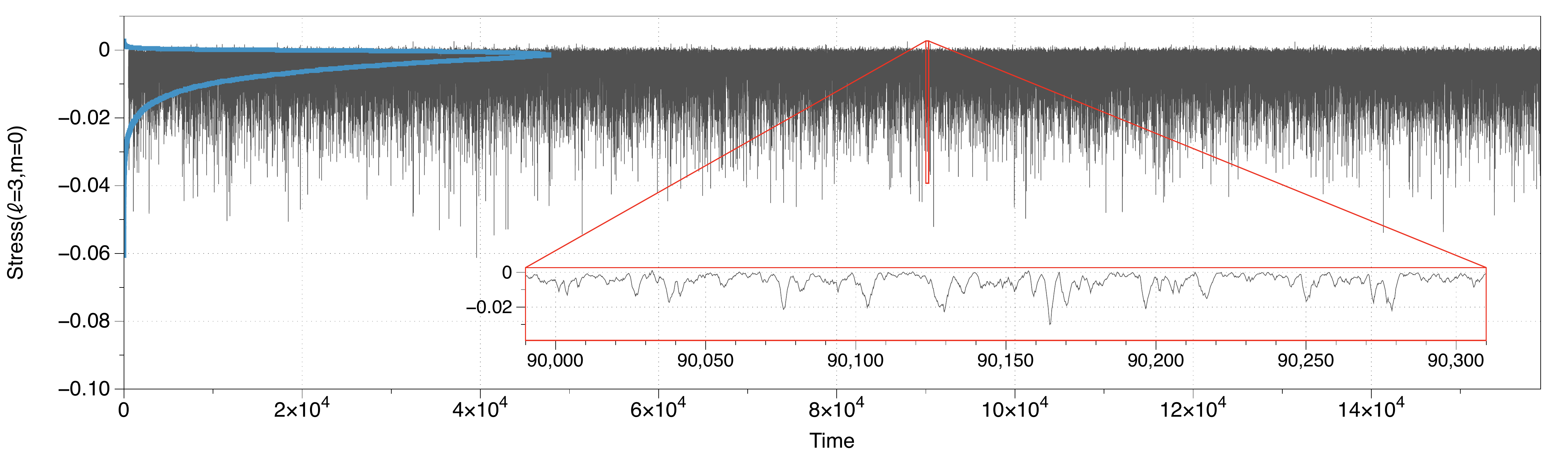}
\includegraphics[width = 1.0\textwidth]{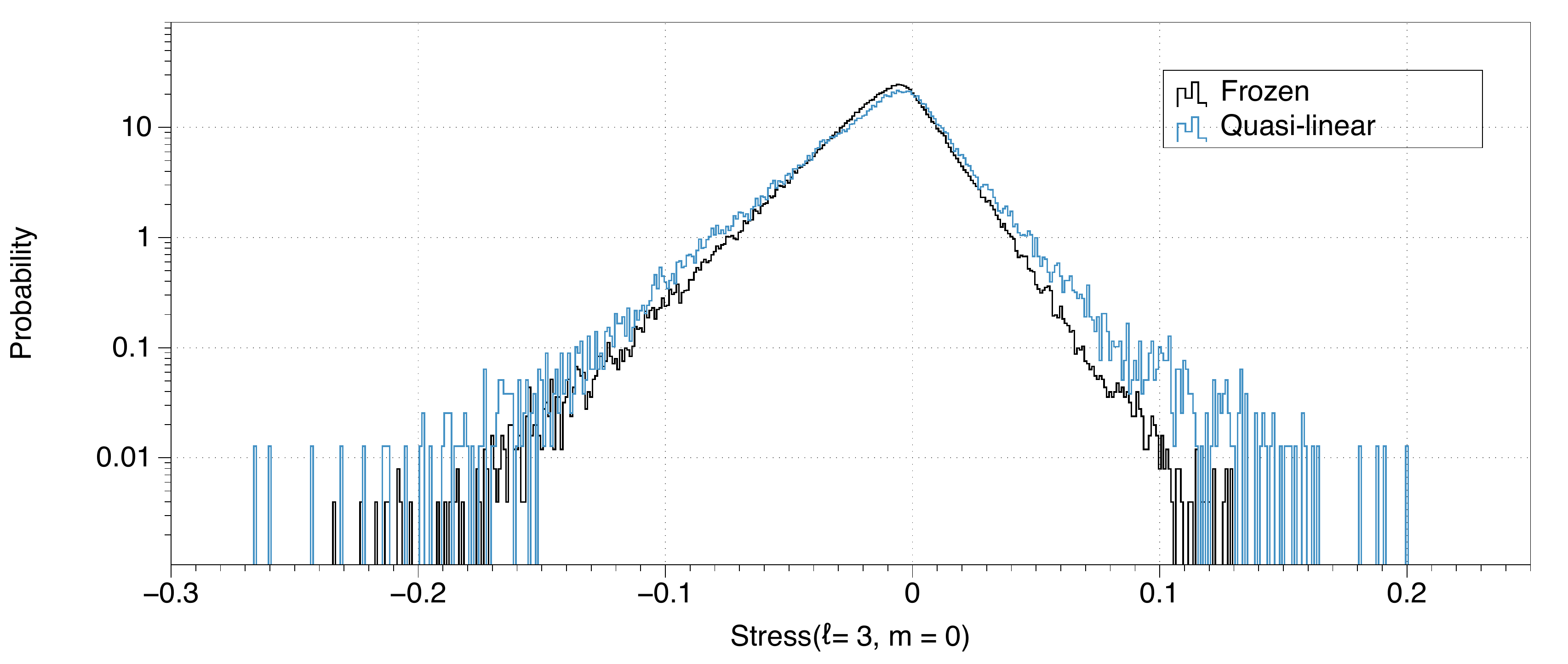}
\caption{Top:  Stress exerted on a component of the mean flow by eddies forced with stochastic Gaussian white noise exhibits large excursions.    Bottom: The invariant measure of the Reynold's stress is highly non-Gaussian and well approximated by two exponentials.  The distribution for the dynamically-evolving mean flow of the quasilinear approximation is, in this case, agrees well with a simulation in which the mean-flow is frozen at its time-mean value.  The generating functional for the statistics can be found by solving the Ricatti equation, a non-linear generalization of the Lyapunov equation, matching simulation well (not shown; see \citep{Bouchet:2018er}).}
\label{fig:Stress}
\end{figure*}

The statistical description of a flow is more than just its low-order moments.  Large Deviation Theory \citep{Touchette:2009eb} provides access to rare events and large departures from the mean.  For example, the statistics of the stress exerted on the mean-flow of a stochastically-driven jet can be recovered analytically, in the limit of a slowly-changing mean-flow, by solving a non-linear version of the Lyapunov equation called the Ricatti equation \citep{Bouchet:2018er}.  Because the equations of motion for the equal-time cumulants close at second order, it may seem surprising that highly non-Gaussian statistics appear, as evident in \textbf{Figure \ref{fig:Stress}}.  This can, however, be understood as a consequence of the averaging operation which is a spatial average in one (zonal) direction and not an ensemble average.  
Large deviation theory has been used to access rare transitions in barotropic flows \citep{Bouchet:2014fm,Laurie:2015co} but it
is currently not known how to apply it to quasilinear dynamics with rapidly fluctuating mean flows (\textbf{Figure \ref{fig:MergingIntermittentJets}}), nor to deterministically driven flows.  This could be a fruitful direction for future research.

\subsection{Dimensional Reduction via Machine Learning}
\label{DimensionalReduction}

Power spectra produced by quasilinear dynamics typically show that fewer wave modes are excited than in fully nonlinear simulations.  This facet has been explored by combining the quasilinear approximation proper orthogonal decomposition (POD; equivalent to principle component analysis or PCA, and empirical orthogonal functions or EOFs) \citep{Allawala:2020kq,Skitka:2020co,Nikolaidis2021}.  {Nowadays POD is often called a form of unsupervised machine learning despite its venerable history that predates machine learning.} 
The reduced dimensionality of the QL approximation can be quantified by POD of the second cumulant.  Quasilinear approximations and second order cumulant expansions of reduced dimensionality may then continue to capture the important features of the flows, with reduced computational work. 

A new basis of lower dimensionality that optimally represents the second cumulant may be found by Schmidt decomposition of the zonally-averaged second moment:  
\begin{eqnarray}
\cb(\bhr_1, \bhr_2) = \sum_i \lambda_i~ {\bf v}_i(\bhr_1) \otimes {\bf v}_i(\bhr_2)\ .
\end{eqnarray}
Here ${\bf v}_i(\bhr)$ is an eigenvector of the second cumulant with eigenvalue $\lambda_i$ that should be both real and non-negative \citep{kraichnan1980realizability}.  Retention of only the eigenvectors with eigenvalues that exceed a preset threshold yields a new basis for reduced QL or CE2. Application of POD to forms of DSS such as CE2 is particularly attractive because low-order statistics are typically much smoother than instantaneous dynamics \citep{Allawala:2020kq} and thus it is natural to work directly with the EOMs for the statistics themselves.  

Areas for future exploration include the use of neural networks such as autoencoders for dimensional reduction \citep{Spears2018} as well as the use of machine learning to adaptively evolve the optimal basis for DSS.

\subsection{Reduced Models for Deeper Understanding}
\label{Reduced}

Perhaps as important as the reduction in computational complexity is the use of reduced statistical models to identify the important modes, leading to physical insight not apparent in the fully nonlinear dynamics.  For example, jet formation as a problem of pattern formation has seen progress in the work of Parker and Krommes \citep{Parker:2013hy,Parker:2014fc} who used the CE2 framework to derive a real Landau-Ginzburg equation for the study of jet bifurcations near the threshold for jet formation.  

A richer system that is still not fully understood is the classic problem of fluid flowing through a pipe with a circular cross section.  A pressure difference between the two ends of the pipe drives a volume flux. As the pressure gradient is increased, so does the difference in velocity between the center of the flow and its edges. The increasing importance of inertial forces relative to viscous forces is described by an increasing Reynolds number.  At a critical value of the Reynolds number the flow undergoes a subcritical transition from a smooth laminar state to intermittent turbulence characterized by puffs of turbulence flow separated by laminar regions.  Further increases of the Reynolds number leads eventually to turbulence throughout the pipe.  Note that all this turbulent dynamics occurs when the laminar profile is linearly stable against small perturbations; laminar flow is only disrupted and driven turbulent by a significant disturbance.

Experiments and simulations now provide strong evidence that the transition to turbulence in pipe flow is a second-order phase transition in the universality class of directed percolation, with the same critical exponents \citep{Shih:2015dl, Barkley:2016eo}.   This naturally raises the question as to whether a map can be found between pipe flow and flow through a porous material.  Two possible routes have been suggested for constructing such a map.  \citep{Barkley:2016eo} summarizes a body of work that exploits a deep connection between pipe flow and excitable and bistable media, and is based on the interplay between the mean shear and turbulence.  \citep{Shih:2015dl} suggest thinking in terms of predator - prey ecology, with the turbulence (prey) generated by an instability acting to drive a zonal flow (predator) that, when large enough, suppresses the turbulence, resulting in Lotka-Volterra type dynamics. Ultimately, these kinds of maps could provide a tool for determining when a flow of any scale transitions from laminar to turbulent, while avoiding computationally intensive simulations.  It would be interesting to determine whether or not either of these simplified models emerge from dimensionally-reduced quasilinear simulations of pipe flow \citep{Willis2007}.

\section{BEYOND QUASILINEARITY}
\label{BeyondQL}
 
Quasilinear theories include a broad range of different approximations.  Averages can be taken over space, time, or ensembles.  The spatial mean itself can be generalized to include long-wavelength fluctuations (GQL). Discarded interactions can be parameterized by the introduction of stochastic forcing. An important lesson that can be drawn from the body of literature is that frequently we learn more about fluids when the quasilinear approximation breaks down, than when it works well.  

\begin{figure*}
\includegraphics[width = 1.0\textwidth]{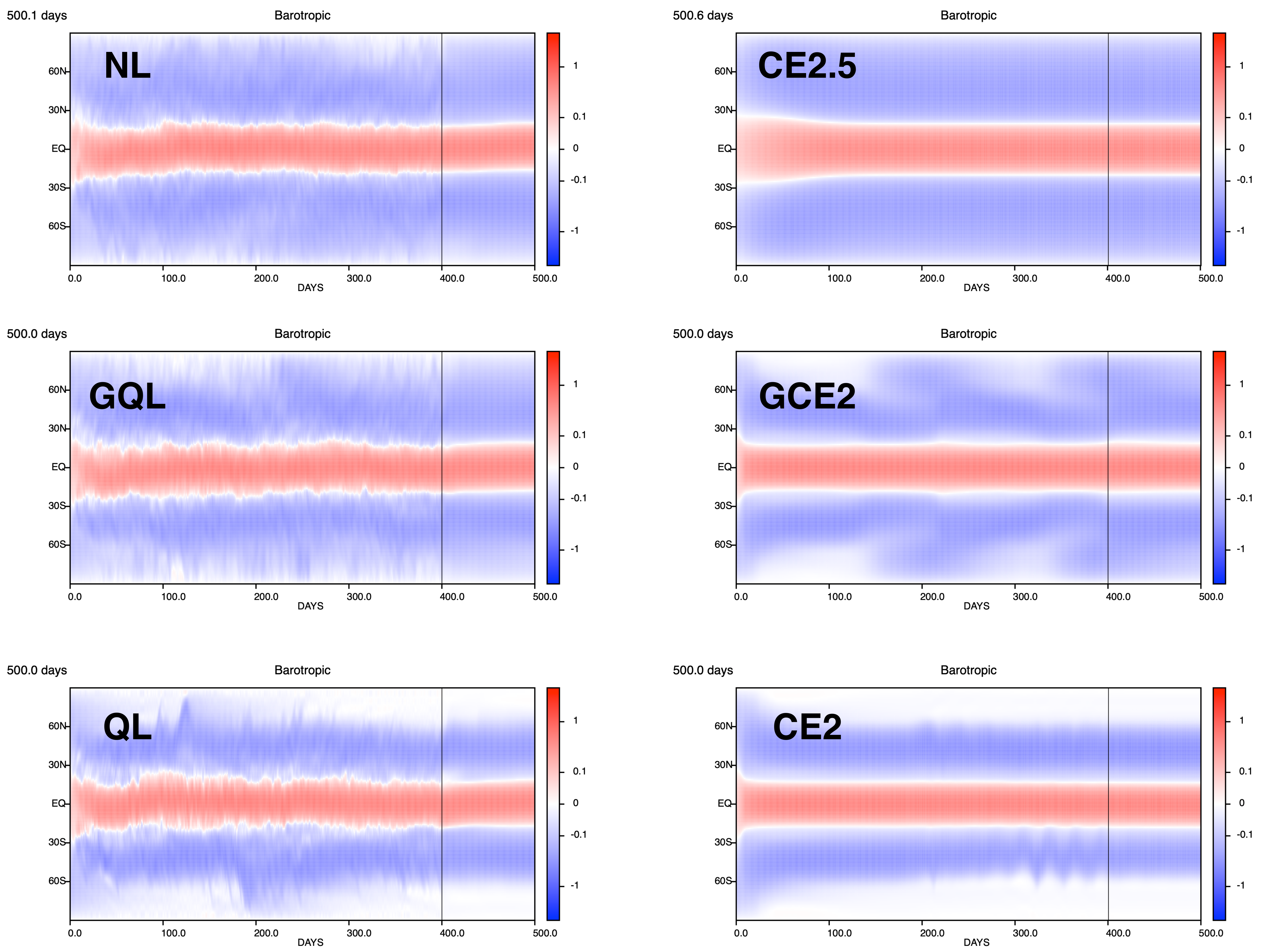}
\caption{Hovm\"oller plots of the zonal mean zonal velocity of a stochastically forced barotropic jet on a rotating sphere. A detailed description of the model can be found in  \citep{marston2014direct}.  Time averaging commences at 400 days as indicated by the black vertical line.  The different approximations correspond to those shown in the schematics of \textbf{Figure \ref{fig:Approximations}}. For CE2.5 the eddy damping time is chosen to be $\tau = 2$ days; for GQL and GCE2 the wavenumber cutoff $\Lambda = 1$, the minimal extension beyond QL and CE2. Note that counterflow is not found at high latitudes in the QL approximation and its closure (CE2) due to the absence of eddy + eddy $\rightarrow$ eddy interactions that are responsible for the transport of angular momentum away from low latitudes.  This defect of the quasilinear approximation is corrected in the GQL, GCE2 and CE2.5 approximations.}
\label{fig:DSS-Comparison}
\end{figure*}

Nevertheless it can also be enlightening to push statistical theories beyond quasilinearity. Extending the cumulant expansion beyond second order is relatively straightforward, and brings in the neglected ``pain in the neck'' eddy + eddy $\rightarrow$ eddy processes neglected at second order.  \textbf{Figure \ref{fig:DSS-Comparison}} shows a comparison of the zonal mean flow generated by rotating barotropic flow on the sphere stirred by stochastic forcing as calculated in the various approximations discussed in this review (see \textbf{Figure \ref{fig:Approximations}}). 

\subsection{Higher-order Cumulant Expansions}
\label{CE3}

Closure at third-order, CE3, is achieved by discarding the contribution to the tendencies of the cumulants from the 4th and higher cumulants.  The approximation is no longer realizable unless further measures are taken \citep{marston2014direct}, but may be made realizable by assuming that the third cumulant evolves rapidly in comparison with the first and second cumulant. This assumption, in combination with the introduction of an eddy-damping time $\tau$ to parameterize the missing 4th cumulant,
means that the prognostic equation for the third cumulant may be promoted to a diagnostic one.  A further simplification that leads to faster computation involves the neglect of all contributions involving the first order cumulant in the equation for the third cumulant. The third order cumulant is then given diagnostically by a product of two second order cumulants coupled together by the quadratic nonlinearity:
\begin{equation}
 \frac{1}{\tau}~ \cb(\bhr_1, \bhr_2, \bhr_3) = \left\lbrace \mcN[\cb(\bhr_1, \bhr_2), \cb(\bhr_1, \bhr_3)] \right\rbrace.
 \label{CE2.5}
\end{equation}
In the cumulant hierarchy, the third cumulant then drives the tendency of the second cumulant much like the covariance matrix of stochastic forcing.  This is the CE2.5 approximation \citep{marston2014direct,Li2021_dyn} and {it is often found to be the case that the statistics are insensitive to the precise value of $\tau$, suggesting that further reductions in computational intensity may be achievable.}  CE2.5 may be regarded as a generalization of the eddy-damped quasi-normal Markovian (EDQNM) approximation \citep{orszag1970,orszag1977} to treat anisotropic and inhomogeneous flows \cite{Legras:1980er}. 

\subsection{Fokker-Planck Equation and Direct Interaction Approximation}
\label{DIA}

Expansions in equal-time cumulants beyond second order require a sharp increase in computational power as the third and higher cumulants involve more and more coordinates and thus have increaingly high dimension.  This ``curse of dimensionality" becomes acute for more ambitious forms of DSS, such as the Fokker-Planck equation that, for continuous fluid systems, takes the form of a \emph{functional} differential equation \citep{Venturi:2018bp}.  The stationary solution of the Fokker-Planck equation is the probability distribution function or invariant measure that includes information about all equal-time correlations, and thus cumulants, as well as rare events.  As it is a linear equation, the stationary state of the Fokker-Planck equation may be found by numerical linear algebra algorithms \citep{Allawala:2016hx}.  A recent effort to reduce the dimensionality of the functional equation may show some promise \citep{Chen:2017ku}.   

By contrast the Direct Interaction Approximation (DIA) \citep{Kraichnan1959b,kraichnan1961,Kraichnan1964a} encodes information about non-equal time correlations but thus far only limited progress has been made on anisotropic \citep{Orszag1987} and inhomogeneous \citep{Okane2004,Frederiksen:2018uc} turbulence, though see also the excellent review by \citep{yokoi19} who also discusses such elaborations as the extension of the DIA approximation to two scales. This nevertheless may be an interesting avenue for future exploration, especially for wave-dominated dynamics for which correlations at different times can be large.

\subsection{Functional Renormalization Group}
\label{frg}

The success of the renormalization group approach to understanding equilibrium critical phenomena led to enthusiastic efforts in the 1980s to try to understand power law scaling seen in fluids experiments and in simulations of homogeneous and isotropic turbulence but these ran into many technical difficulties. The functional renormalization-group approach \citep{DUPUIS20211} that generalizes ordinary renormalization group calculations to track the flow of entire functionals of the fields appearing in the effective action, instead of a small set of coupling constants, shows promise in surmounting some of these difficulties, going beyond K41 theory \citep{yag_1994} to describe intermittency \citep{Canet:2016jl}. The approach not only quantifies how energy cascades from lengthscale to lengthscale, as predicted by Kolmogorov scaling, but may also describe physics that goes beyond Kolmogorov theory. Most effort has been focused on the problem of homogeneous and isotropic turbulence for which the correlations have the highest symmetry, but it appears that the approach may be amenable to treating inhomogeneous and anisotropic turbulence.  The overview presented here suggests that technical difficulties encountered due to the reduced symmetry of such flows may be compensated by the existence of systematic expansion parameters that could be exploited.

\section{CONCLUSION}
\label{conclusion}

This review has summarised historical and current efforts to provide a theoretical and computational framework for inhomogeneous and anisotropic turbulence interacting with mean flows or magnetic fields. The underlying interactions for such systems differ markedly from the oft-studied, but less oft-observed, paradigm case of homogeneous, anisotropic flows --- and the methods employed are likewise qualitatively different.  Many open questions remain and further progress is anticipated. 

We have discussed how quasilinear models (and generalisations thereof) can yield a first-order approximation for the relevant dynamics and may also lead to the development of statistical theories for the evolution of the low-order statistics. Such theories may be extended to treat nonlinearities more fully via the inclusion of more of the triadic interactions (GQL and GCE2) or higher cumulants (CE3 / CE2.5).  These extensions rapidly becomes computationally expensive owing to the curse of dimensionality and the goal then becomes one of constructing computationally efficient algorithms via model reduction.  Some early progress along this line was sketched.  

We predict that future research will involve the incorporation of data-driven methods into this statistical framework --- for example via learning the low-order statistics from local models to enable the efficient integration of global models. We also believe that the local nature of the statistical framework will naturally lend itself to GPU, and possibly quantum, computation.

\section*{DISCLOSURE STATEMENT}
The authors are not aware of any affiliations, memberships, funding, or financial holdings that
might be perceived as affecting the objectivity of this review. 

\section*{ACKNOWLEDGEMENTS}
We thank Freddy Bouchet, Greg Chini, Baylor Fox-Kemper, Jack Herring, Rich Kerswell, Paul Kushner, Kuan Li, Girish Nivarti, Jeff Oishi, Tapio Schneider and Eli Tziperman for helpful discussions. The authors are supported in part by a grant from the Simons Foundation (Grant No. 662962, GF). SMT would also like to acknowledge support of funding from the European Union Horizon 2020 research and innovation programme (grant agreement no. D5S-DLV-786780).

\section{TERMS AND DEFINITIONS}
\label{terms}

\begin{enumerate}
\item CE2: Second-order cumulant expansion.
\item {CE2.5: Closure with third cumulant determined diagnostically from second cumulants.}
\item CE3: Third-order cumulant expansion.
\item DIA: Direct interaction approximation.
\item DNS: Direct numerical simulation.
\item DSS: Direct statistical simulation. 
\item EDQNM: Eddy-damped quasi-normal Markovian approximation. 
\item EOM: Equation of motion.
\item GCE2: Generalized second order cumulant expansion.
\item GQL: Generalized quasilinear.
\item NL: (Fully) Nonlinear.
\item POD: Proper Orthogonal Decomposition
\item QL: Quasilinear.
\item RDT: Rapid distortion theory.
\item RNL: Restricted non-linear approximation.
\item S3T / SSST: Stochastic structural stability theory. 
\item SSD: Statistical state dynamics.
\end{enumerate}


\begin{thebibliography}{}
\expandafter\ifx\csname natexlab\endcsname\relax\def\natexlab#1{#1}\fi

\bibitem[Ait-Chaalal et~al.(2016)Ait-Chaalal, Schneider, Meyer \&
  Marston]{ait2016cumulant}
Ait-Chaalal F, Schneider T, Meyer B, Marston J. 2016.
{Cumulant expansions for atmospheric flows}.
\textit{New Journal of Physics} 18(2):025019

\bibitem[Allawala \& Marston(2016)]{Allawala:2016hx}
Allawala A, Marston JB. 2016.
{Statistics of the stochastically-forced Lorenz attractor by the Fokker-Planck
  equation and cumulant expansions}.
\textit{Physical Review E} nlin.CD(5):052218

\bibitem[Allawala et~al.(2020)Allawala, Tobias \& Marston]{Allawala:2020kq}
Allawala A, Tobias SM, Marston JB. 2020.
{Dimensional reduction of direct statistical simulation}.
\textit{Journal of Fluid Mechanics} 898:533 -- 18

\bibitem[Arakawa(1966)]{arakawa1966}
Arakawa A. 1966.
{Computational design for long-term numerical integration of the equations of
  fluid motion: Two-dimensional incompressible flow. Part I}.
\textit{J. Comp. Phys.} 1:119--143

\bibitem[Bakas \& Ioannou(2011)]{bakas2011structural}
Bakas NA, Ioannou PJ. 2011.
{Structural stability theory of two-dimensional fluid flow under stochastic
  forcing}.
\textit{Journal of Fluid Mechanics} 682:332--361

\bibitem[Bakas \& Ioannou(2013)]{bakas2013emergence}
Bakas NA, Ioannou PJ. 2013.
{Emergence of large scale structure in barotropic $\beta$-plane turbulence}.
\textit{Physical review letters} 110(22):224501

\bibitem[Bakas \& Ioannou(2014)]{bakas2014theory}
Bakas NA, Ioannou PJ. 2014.
{A theory for the emergence of coherent structures in beta-plane turbulence}.
\textit{Journal of Fluid Mechanics} 740:312--341

\bibitem[Barkley(2006)]{Barkley_2006}
Barkley D. 2006.
Linear analysis of the cylinder wake mean flow.
\textit{Europhysics Letters ({EPL})} 75(5):750--756

\bibitem[Barkley(2016)]{Barkley:2016eo}
Barkley D. 2016.
{Theoretical perspective on the route to turbulence in a pipe}.
\textit{Journal of Fluid Mechanics} 803:15 -- 80

\bibitem[Batchelor(1953)]{batchelor1953}
Batchelor GK. 1953.
The theory of homogeneous turbulence.
Cambridge university press

\bibitem[Batchelor \& Proudman(1954)]{Batchelor1954}
Batchelor GK, Proudman I. 1954.
{The Effect of Rapid Distortion of a Fluid in Turbulent Motion}.
\textit{The Quarterly Journal of Mechanics and Applied Mathematics}
  7(1):83--103

\bibitem[Bengana \& Tuckerman(2021)]{bt_2021}
Bengana Y, Tuckerman LS. 2021.
Frequency prediction from exact or self-consistent mean flows.
\textit{Phys. Rev. Fluids} 6(6):063901

\bibitem[Bouchet et~al.(2014)Bouchet, Laurie \& Zaboronski]{Bouchet:2014fm}
Bouchet F, Laurie J, Zaboronski O. 2014.
{Langevin Dynamics, Large Deviations and Instantons for the Quasi-Geostrophic
  Model and Two-Dimensional Euler Equations}.
\textit{Journal of Statistical Physics} 156(6):1066 -- 1092

\bibitem[Bouchet et~al.(2018)Bouchet, Marston \& Tangarife]{Bouchet:2018er}
Bouchet F, Marston JB, Tangarife T. 2018.
{Fluctuations and large deviations of Reynolds stresses in zonal jet dynamics}.
\textit{Physics of Fluids} 30(1):015110--20

\bibitem[Bouchet et~al.(2013)Bouchet, Nardini \&
  Tangarife]{bouchetnardinietal2013}
Bouchet F, Nardini C, Tangarife T. 2013.
Kinetic theory of jet dynamics in the stochastic barotropic and 2d
  navier-stokes equations.
\textit{Journal of Statistical Physics} 153(4):572--625

\bibitem[Bretheim et~al.(2015)Bretheim, Meneveau \& Gayme]{bmg_2015}
Bretheim JU, Meneveau C, Gayme DF. 2015.
Standard logarithmic mean velocity distribution in a band-limited restricted
  nonlinear model of turbulent flow in a half-channel.
\textit{Physics of Fluids} 27(1):011702

\bibitem[Canet et~al.(2016)Canet, Delamotte \& Wschebor]{Canet:2016jl}
Canet L, Delamotte B, Wschebor N. 2016.
{Fully developed isotropic turbulence: Nonperturbative renormalization group
  formalism and fixed-point solution}.
\textit{Physical Review E} 93(6):299 -- 26

\bibitem[Chen \& Majda(2017)]{Chen:2017ku}
Chen N, Majda AJ. 2017.
{Beating the curse of dimension with accurate statistics for the Fokker-Planck
  equation in complex turbulent systems.}
\textit{Proceedings of the National Academy of Sciences} 114(49):12864 -- 12869

\bibitem[{Child} et~al.(2016){Child}, {Hollerbach}, {Marston} \&
  {Tobias}]{chmt2016}
{Child} A, {Hollerbach} R, {Marston} B, {Tobias} S. 2016.
{Generalised quasilinear approximation of the helical magnetorotational
  instability}.
\textit{Journal of Plasma Physics} 82(3):905820302

\bibitem[Chini et~al.(2014)Chini, Malecha \& Dreeben]{cmd2014}
Chini GP, Malecha Z, Dreeben TD. 2014.
Large-amplitude acoustic streaming.
\textit{Journal of Fluid Mechanics} 744:329--351

\bibitem[Chini et~al.(2022)Chini, Michel, Julien, Rocha \&
  Caulfield]{cmjrc_2022}
Chini GP, Michel G, Julien K, Rocha CB, Caulfield CcP. 2022.
Exploiting self-organized criticality in strongly stratified turbulence.
\textit{Journal of Fluid Mechanics} 933:A22

\bibitem[Connaughton et~al.(2015)Connaughton, Nazarenko \&
  Quinn]{Connaughton2015}
Connaughton C, Nazarenko S, Quinn B. 2015.
{Rossby and drift wave turbulence and zonal flows: The Charney--Hasegawa--Mima
  model and its extensions}.
\textit{Physics Reports} 604:1--71

\bibitem[Constantinou et~al.(2014{\natexlab{a}})Constantinou, Farrell \&
  Ioannou]{Constantinou:2013fh}
Constantinou NC, Farrell BF, Ioannou PJ. 2014{\natexlab{a}}.
{Emergence and equilibration of jets in beta-plane turbulence: applications of
  Stochastic Structural Stability Theory}.
\textit{Journal of the Atmospheric Sciences} 71:1818 --- 1842

\bibitem[Constantinou et~al.(2014{\natexlab{b}})Constantinou, Farrell \&
  Ioannou]{constantinoufarrelletal2013}
Constantinou NC, Farrell BF, Ioannou PJ. 2014{\natexlab{b}}.
Emergence and equilibration of jets in beta-plane turbulence: applications of
  stochastic structural stability theory.
\textit{J. Atmos. Sci.} 71:1818--1842

\bibitem[Constantinou et~al.(2016)Constantinou, Farrell \&
  Ioannou]{Constantinou:2016fp}
Constantinou NC, Farrell BF, Ioannou PJ. 2016.
{Statistical State Dynamics of Jet--Wave Coexistence in Barotropic Beta-Plane
  Turbulence}.
\textit{Journal of Atmospheric Sciences} 73(5):2229 -- 2253

\bibitem[Davidson(2015)]{davidson2015turbulence}
Davidson PA. 2015.
Turbulence: an introduction for scientists and engineers.
Oxford university press

\bibitem[Delplace et~al.(2017)Delplace, Marston \& Venaille]{Delplace:2017kq}
Delplace P, Marston JB, Venaille A. 2017.
{Topological origin of equatorial waves}.
\textit{Science} 358(6366):1075--1077

\bibitem[DelSole \& Farrell(1996)]{delsolefarrell1996}
DelSole T, Farrell BF. 1996.
The quasi-linear equilibration of a thermally maintained, stochastically
  excited jet in a quasigeostrophic model.
\textit{Journal of the atmospheric sciences} 53(13):1781--1797

\bibitem[Domaradzki \& Orszag(1987-09)]{Orszag1987}
Domaradzki JA, Orszag SA. 1987-09.
{Numerical solutions of the direct interaction approximation equations for
  anisotropic turbulence}.
\textit{Journal of Scientific Computing (ISSN 0885-7474)} 2:227 -- 248

\bibitem[Dupuis et~al.(2021)Dupuis, Canet, Eichhorn, Metzner, Pawlowski
  et~al.]{DUPUIS20211}
Dupuis N, Canet L, Eichhorn A, Metzner W, Pawlowski J, et~al. 2021.
The nonperturbative functional renormalization group and its applications.
\textit{Physics Reports} 910:1--114

\bibitem[Eyink \& Sreenivasan(2006)]{es_2006}
Eyink GL, Sreenivasan KR. 2006.
Onsager and the theory of hydrodynamic turbulence.
\textit{Rev. Mod. Phys.} 78(1):87--135

\bibitem[Falkovich et~al.(2001)Falkovich, Gaw\ifmmode~\mbox{\c{e}}\else
  \c{e}\fi{}dzki \& Vergassola]{fal_etal_2001}
Falkovich G, Gaw\ifmmode~\mbox{\c{e}}\else \c{e}\fi{}dzki K, Vergassola M.
  2001.
Particles and fields in fluid turbulence.
\textit{Rev. Mod. Phys.} 73(4):913--975

\bibitem[Farrell \& Ioannou(1993)]{Farrell1993StochasticFO}
Farrell BF, Ioannou PJ. 1993.
Stochastic forcing of the linearized navier--stokes equations.
\textit{Physics of Fluids} 5:2600--2609

\bibitem[Farrell \& Ioannou(2003)]{farrellioannou2003}
Farrell BF, Ioannou PJ. 2003.
Structural stability of turbulent jets.
\textit{Journal of the Atmospheric Sciences} 60:2101--2118

\bibitem[Farrell \& Ioannou(2007)]{farrellioannou2007}
Farrell BF, Ioannou PJ. 2007.
Structure and spacing of jets in barotropic turbulence.
\textit{J. Atmos. Sci.} 64:3652--3665

\bibitem[Farrell et~al.(2016)Farrell, Ioannou, Jim{\'e}nez, Constantinou,
  Lozano-Dur{\'a}n \& Nikolaidis]{farrell_etal_2016}
Farrell BF, Ioannou PJ, Jim{\'e}nez J, Constantinou NC, Lozano-Dur{\'a}n A,
  Nikolaidis MA. 2016.
A statistical state dynamics-based study of the structure and mechanism of
  large-scale motions in plane poiseuille flow.
\textit{Journal of Fluid Mechanics} 809:290--315

\bibitem[{Feynman} et~al.(1964){Feynman}, {Leighton}, {Sands} \&
  {Treiman}]{feyn_etal_1964}
{Feynman} RP, {Leighton} RB, {Sands} M, {Treiman} SB. 1964.
{The Feynman Lectures on Physics}.
\textit{Physics Today} 17(8):45

\bibitem[Frederiksen \& Okane(2018)]{Frederiksen:2018uc}
Frederiksen JS, Okane TJ. 2018.
{Markovian inhomogeneous closures for Rossby waves and turbulence over
  topography}.
\textit{Journal of Fluid Mechanics} 858:45--70

\bibitem[Fried et~al.(1960)Fried, Gell-Mann, Jackson \& Wyld]{Fried1960}
Fried BD, Gell-Mann M, Jackson JD, Wyld HW. 1960.
{Longitudinal plasma oscillations in an electric field}.
\textit{Journal of Nuclear Energy. Part C, Plasma Physics, Accelerators,
  Thermonuclear Research} 1(4):190

\bibitem[Frisch(1995)]{frisch1995turbulence}
Frisch U. 1995.
{Turbulence: the legacy of AN Kolmogorov}.
Cambridge university press

\bibitem[Frishman(2017)]{Frishman:2017}
Frishman A. 2017.
{The culmination of an inverse cascade: Mean flow and fluctuations}.
\textit{Physics of Fluids} 29(12):125102

\bibitem[Frishman \& Herbert(2018)]{Frishman:2018}
Frishman A, Herbert C. 2018.
{Turbulence Statistics in a Two-Dimensional Vortex Condensate}.
\textit{Physical Review Letters} 120(20):204505

\bibitem[Hasselmann(1966)]{Hasselmann:1966hj}
Hasselmann K. 1966.
{Feynman diagrams and interaction rules of wave - wave scattering processes}.
\textit{Reviews of Geophysics} 4(1):1 -- 32

\bibitem[Held \& Phillips(1987)]{HELD:1987vj}
Held IM, Phillips PJ. 1987.
{Linear and nonlinear barotropic decay on the sphere}.
\textit{Journal of the Atmospheric Sciences} 44(1):200 -- 207

\bibitem[Hern{\'a}ndez et~al.(2022{\natexlab{a}})Hern{\'a}ndez, Yang \&
  Hwang]{hern_2022a}
Hern{\'a}ndez CG, Yang Q, Hwang Y. 2022{\natexlab{a}}.
Generalised quasilinear approximations of turbulent channel flow. part 1.
  streamwise nonlinear energy transfer.
\textit{Journal of Fluid Mechanics} 936:A33

\bibitem[Hern{\'a}ndez et~al.(2022{\natexlab{b}})Hern{\'a}ndez, Yang \&
  Hwang]{hern2022}
Hern{\'a}ndez CG, Yang Q, Hwang Y. 2022{\natexlab{b}}.
Generalised quasilinear approximations of turbulent channel flow: Part 2.
  spanwise scale interactions.
ArXiv:2112.01972

\bibitem[Herring(1963)]{Herring:1963}
Herring JR. 1963.
{Investigation of Problems in Thermal Convection}.
\textit{Journal of the Atmospheric Sciences} 20(4):325 -- 338

\bibitem[Hunt \& Carruthers(1990)]{Hunt1990}
Hunt JCR, Carruthers DJ. 1990.
{Rapid distortion theory and the `problems' of turbulence}.
\textit{Journal of Fluid Mechanics} 212(-1):497--532

\bibitem[Hwang \& Cossu(2010{\natexlab{a}})]{Hwang2010AmplificationOC}
Hwang Y, Cossu C. 2010{\natexlab{a}}.
Amplification of coherent streaks in the turbulent couette flow: an
  input--output analysis at low reynolds number.
\textit{Journal of Fluid Mechanics} 643:333 -- 348

\bibitem[Hwang \& Cossu(2010{\natexlab{b}})]{hwang_cossu_2010}
Hwang Y, Cossu C. 2010{\natexlab{b}}.
Linear non-normal energy amplification of harmonic and stochastic forcing in
  the turbulent channel flow.
\textit{Journal of Fluid Mechanics} 664:51--73

\bibitem[Kaspi et~al.(2020)Kaspi, Galanti, Showman, Stevenson, Guillot
  et~al.]{kaspi2020}
Kaspi Y, Galanti E, Showman AP, Stevenson DJ, Guillot T, et~al. 2020.
{Comparison of the Deep Atmospheric Dynamics of Jupiter and Saturn in Light of
  the Juno and Cassini Gravity Measurements}.
\textit{Space Science Reviews} 216(5):84

\bibitem[Kellam(2019)]{kellam2019}
Kellam C. 2019.
Generalized quasilinear simulation of turbulent channel flow.
Ph.D. thesis, University of New Hampshire, Durham, New Hampshire

\bibitem[Kraichnan(1959)]{Kraichnan1959b}
Kraichnan RH. 1959.
The structure of isotropic turbulence at very high {R}eynolds numbers.
\textit{J. Fluid Mech.} 5:497--543

\bibitem[Kraichnan(1961)]{kraichnan1961}
Kraichnan RH. 1961.
Dynamics of nonlinear stochastic systems.
\textit{J. Math.\ Phys.} 2:124--148Erratum: J. Math.\ Phys.\ {\bf 3}, 205
  (1962)

\bibitem[Kraichnan(1964)]{Kraichnan1964a}
Kraichnan RH. 1964.
Decay of isotropic turbulence in the direct-interaction approximation.
\textit{Phys.\ Fluids} 7:1030--1048

\bibitem[Kraichnan(1980)]{kraichnan1980realizability}
Kraichnan RH. 1980.
{Realizability Inequalities and Closed Moment Equations}.
\textit{Annals of the New York Academy of Sciences} 357(1):37--46

\bibitem[Kraichnan(1985)]{kraichnan1985}
Kraichnan RH. 1985.
{Decimated amplitude equations in turbulence dynamics}. In \textit{Theoretical
  Approaches to Turbulence}, eds. DL~Dwoyer, MY~Hussaini, RG~Voight. New York:
  Springer,  91--135

\bibitem[{Krause} \& {Raedler}(1980)]{krauraed:1980}
{Krause} F, {Raedler} K. 1980.
{Mean-field magnetohydrodynamics and dynamo theory}.
Oxford, Pergamon Press, Ltd., 1980.~271 p.

\bibitem[Laurie et~al.(2014)Laurie, Boffetta, Falkovich, Kolokolov \&
  Lebedev]{Laurie:2014dn}
Laurie J, Boffetta G, Falkovich G, Kolokolov I, Lebedev V. 2014.
{Universal Profile of the Vortex Condensate in Two-Dimensional Turbulence}.
\textit{Physical Review Letters} 113(25):254503 -- 5

\bibitem[Laurie \& Bouchet(2015)]{Laurie:2015co}
Laurie J, Bouchet F. 2015.
{Computation of rare transitions in the barotropic quasi-geostrophic
  equations}.
\textit{New Journal of Physics} 17(1):1 -- 25

\bibitem[Ledoux et~al.(1961)Ledoux, Schwarzchild \& Spiegel]{Ledoux:1961ci}
Ledoux P, Schwarzchild M, Spiegel EA. 1961.
{On the Spectrum of Turbulent Convection}.
\textit{Astrophysical Journal} 133(1):184 -- 197

\bibitem[Legras(1980-01)]{Legras:1980er}
Legras B. 1980-01.
{Turbulent phase shift of Rossby waves}.
\textit{Geophysical \& Astrophysical Fluid Dynamics} 15(1):253 -- 281

\bibitem[Li et~al.(2021{\natexlab{a}})Li, Marston, Saxena \&
  Tobias]{li2021_l63}
Li K, Marston J, Saxena S, Tobias SM. 2021{\natexlab{a}}.
Direct statistical simulation of the lorenz63 system.
\textit{Chaos} 32:043111

\bibitem[Li et~al.(2021{\natexlab{b}})Li, Marston \& Tobias]{Li2021_dyn}
Li K, Marston JB, Tobias SM. 2021{\natexlab{b}}.
Direct statistical simulation of low-order dynamo systems.
\textit{Proceedings of the Royal Society A} 477

\bibitem[Malkus(1954)]{Malkus:1954dh}
Malkus WVR. 1954.
{The Heat Transport and Spectrum of Thermal Turbulence}.
\textit{Proceedings of the Royal Society of London. Series A, Mathematical and
  Physical Sciences} 225(1161):196 -- 212

\bibitem[Mantic-Lugo et~al.(2015)Mantic-Lugo, Arratia \&
  Gallaire]{Mantic-Lugo:2014}
Mantic-Lugo V, Arratia C, Gallaire F. 2015.
A self-consistent model for the saturation dynamics of the vortex shedding
  around the mean flow in the unstable cylinder wake.
\textit{Physics Of Fluids} 27(7):19. 074103

\bibitem[Markeviciute \& Kerswell(2022)]{mk_2022}
Markeviciute VK, Kerswell RR. 2022.
Improved assessment of the statistical stability of turbulent flows using
  extended orr-sommerfeld stability analysis.
ArXiv:2201.01540

\bibitem[Marston et~al.(2008)Marston, Conover \& Schneider]{marstonetal2008}
Marston J, Conover E, Schneider T. 2008.
{Statistics of an unstable barotropic jet from a cumulant expansion}.
\textit{Journal of the Atmospheric Sciences} 65(6):1955--1966

\bibitem[{Marston}(2010)]{marston2010}
{Marston} JB. 2010.
{Statistics of the general circulation from cumulant expansions}.
\textit{Chaos} 20(4):041107

\bibitem[Marston(2012)]{marston2012}
Marston JB. 2012.
Planetary atmospheres as nonequilibrium condensed matter.
\textit{Annu.\ Rev.\ Condensed Matter Phys.} 3:285--310

\bibitem[{Marston} et~al.(2016){Marston}, {Chini} \& {Tobias}]{mct2016}
{Marston} JB, {Chini} GP, {Tobias} SM. 2016.
{Generalized Quasilinear Approximation: Application to Zonal Jets}.
\textit{Physical Review Letters} 116(21):214501

\bibitem[{Marston} et~al.(2008){Marston}, {Conover} \&
  {Schneider}]{marstonconoveretal2008}
{Marston} JB, {Conover} E, {Schneider} T. 2008.
{Statistics of an Unstable Barotropic Jet from a Cumulant Expansion}.
\textit{Journal of Atmospheric Sciences} 65:1955

\bibitem[Marston et~al.(2019)Marston, Qi \& Tobias]{marston2014direct}
Marston JB, Qi W, Tobias SM. 2019.
{Direct Statistical Simulation of a jet}. In \textit{{Zonal jets:
  Phenomenology, Genesis and Physics} (arXiv:1412.0381)}, eds. B~Galerpin,
  PL~Read. Cambridge University Press,  332--346.
ArXiv:1412.0381 [physics.flu-dyn]

\bibitem[McKeon \& Sharma(2010)]{mckeonsharma_2010}
McKeon BJ, Sharma AS. 2010.
A critical-layer framework for turbulent pipe flow.
\textit{Journal of Fluid Mechanics} 658:336--382

\bibitem[Meliga(2017)]{meliga_2017}
Meliga P. 2017.
Harmonics generation and the mechanics of saturation in flow over an open
  cavity: a second-order self-consistent description.
\textit{Journal of Fluid Mechanics} 826:503--521

\bibitem[Michel \& Chini(2019)]{mc2019}
Michel G, Chini GP. 2019.
Multiple scales analysis of slow--fast quasi-linear systems.
\textit{Proceedings of the Royal Society A} 475

\bibitem[Moffatt \& Dormy(2019)]{MoffattDormy:2019}
Moffatt H, Dormy E. 2019.
Self-exciting fluid dynamos.
Cambridge University Press, Cambridge

\bibitem[Nikolaidis et~al.(2021)Nikolaidis, Ioannou, Farrell \&
  Lozano-Dur{\'a}n]{Nikolaidis2021}
Nikolaidis MA, Ioannou PJ, Farrell BF, Lozano-Dur{\'a}n A. 2021.
{POD-based study of structure and dynamics in turbulent plane Poiseuille flow:
  comparing quasi-linear simulations to DNS}.
\textit{arXiv:2109.02465}

\bibitem[Nivarti et~al.(2022)Nivarti, Kerswell, Marston \& Tobias]{Nivarti2022}
Nivarti GV, Kerswell RR, Marston JB, Tobias SM. 2022.
Non-equivalence of quasilinear dynamical systems and their statistical
  closures.
\textit{arXiv} ArXiv:2202.04127

\bibitem[Noerdlinger(1963)]{Noerdlinger1963}
Noerdlinger PD. 1963.
{Quasi-Linear Theory of Plasma Oscillations in an Electric Field}.
\textit{Physics of Fluids} 6(8):1196

\bibitem[O'Gorman \& Schneider(2007)]{gormanschneider2007}
O'Gorman PA, Schneider T. 2007.
Recovery of atmospheric flow statistics in a general circulation model without
  nonlinear eddy-eddy interactions.
\textit{Geophysical Research Letters} 34(22):n/a--n/a

\bibitem[Okane \& Frederiksen(2004)]{Okane2004}
Okane TJ, Frederiksen JS. 2004.
{The QDIA and regularized QDIA closures for inhomogeneous turbulence over
  topography}.
\textit{J. Fluid Mech.} 504:133 -- 165

\bibitem[Orszag(1970)]{orszag1970}
Orszag SA. 1970.
Analytical theories of turbulence.
\textit{J. Fluid Mech.} 41:363--386

\bibitem[Orszag(1977)]{orszag1977}
Orszag SA. 1977.
{Lectures on the Statistical Theory of Turbulence}.
\textit{Fluid Dyanmics, Les Houches 1973}

\bibitem[Parker(2021)]{Parker:2020hf}
Parker JB. 2021.
{Topological phase in plasma physics}.
\textit{Journal of Plasma Physics} 87(2):835870202

\bibitem[Parker \& Krommes(2013)]{Parker:2013hy}
Parker JB, Krommes JA. 2013.
{Zonal flow as pattern formation}.
\textit{Physics of Plasmas} 20(10):100703

\bibitem[Parker \& Krommes(2014)]{Parker:2014fc}
Parker JB, Krommes JA. 2014.
{Generation of zonal flows through symmetry breaking of statistical
  homogeneity}.
\textit{New Journal of Physics} :1 -- 29

\bibitem[Parker et~al.(2020)Parker, Marston, Tobias \& Zhu]{pmtz2020}
Parker JB, Marston JB, Tobias SM, Zhu Z. 2020.
Topological gaseous plasmon polariton in realistic plasma.
\textit{Phys. Rev. Lett.} 124(19):195001

\bibitem[Pausch et~al.(2019)Pausch, Yang, Hwang \& Eckhardt]{Pausch2019}
Pausch M, Yang Q, Hwang Y, Eckhardt B. 2019.
{Quasilinear approximation for exact coherent states in parallel shear flows}.
\textit{Fluid Dynamics Research} 51(1):011402

\bibitem[{Pausch} et~al.(2019){Pausch}, {Yang}, {Hwang} \&
  {Eckhardt}]{pausch_2019}
{Pausch} M, {Yang} Q, {Hwang} Y, {Eckhardt} B. 2019.
{Quasilinear approximation for exact coherent states in parallel shear flows}.
\textit{Fluid Dynamics Research} 51(1):011402

\bibitem[Plumb(1977)]{plumb77}
Plumb RA. 1977.
The interaction of two internal waves with the mean flow: Implications for the
  theory of the quasi-biennial oscillation.
\textit{Journal of Atmospheric Sciences} 34(12):1847 -- 1858

\bibitem[Plumley et~al.(2018)Plumley, Calkins, Julien \&
  Tobias]{plumleyetal2018}
Plumley M, Calkins MA, Julien K, Tobias SM. 2018.
Self-consistent single mode investigations of the quasi-geostrophic
  convection-driven dynamo model.
\textit{Journal of Plasma Physics} 84(4):735840406

\bibitem[Plummer et~al.(2019)Plummer, Marston \& Tobias]{pmt2019}
Plummer A, Marston JB, Tobias SM. 2019.
Joint instability and abrupt nonlinear transitions in a differentially rotating
  plasma.
\textit{Journal of Plasma Physics} 85(1):905850113

\bibitem[Pope(2000)]{pope_2000}
Pope SB. 2000.
Turbulent flows.
Cambridge University Press

\bibitem[Saad(2003)]{saad_2003}
Saad Y. 2003.
Iterative methods for sparse linear systems.
Society for Industrial and Applied Mathematics, 2nd ed.

\bibitem[Scott \& Dritschel(2012)]{scottdritschel2012}
Scott RK, Dritschel DG. 2012.
The structure of zonal jets in geostrophic turbulence.
\textit{J. Fluid Mech.} 711:576--598

\bibitem[Shepherd(1990)]{Shepherd:1990bt}
Shepherd TG. 1990.
{Symmetries, Conservation Laws, and Hamiltonian Structure in Geophysical Fluid
  Dynamics}. vol.~32 of \textit{Advances in Geophysics Volume 32}. Elsevier,
  287 -- 338

\bibitem[Shih et~al.(2015)Shih, Hsieh \& Goldenfeld]{Shih:2015dl}
Shih HY, Hsieh TL, Goldenfeld N. 2015.
{Ecological collapse and the emergence of travelling waves at the onset of
  shear turbulence}.
\textit{Nature Physics} 12(3):245 -- 248

\bibitem[Skitka et~al.(2020-03)Skitka, Marston \& Fox-Kemper]{Skitka:2020co}
Skitka J, Marston JB, Fox-Kemper B. 2020-03.
{Reduced-Order Quasilinear Model of Ocean Boundary-Layer Turbulence}.
\textit{Journal of Physical Oceanography} 50(3):537 -- 558

\bibitem[Spears et~al.(2018)Spears, Brase, Bremer, Chen, Field
  et~al.]{Spears2018}
Spears BK, Brase J, Bremer PT, Chen B, Field J, et~al. 2018.
Deep learning: A guide for practitioners in the physical sciences.
\textit{Physics of Plasmas} 25(8):080901

\bibitem[Spiegel(1962)]{Spiegel1962}
Spiegel EA. 1962.
{Thermal turbulence at very small Prandtl number}.
\textit{Journal of Geophysical Research} 67(8):3063--3070

\bibitem[{Squire} \& {Bhattacharjee}(2015)]{sb2015}
{Squire} J, {Bhattacharjee} A. 2015.
{Statistical Simulation of the Magnetorotational Dynamo}.
\textit{Physical Review Letters} 114:085002

\bibitem[Srinivasan \& Young(2012)]{srinivasanyoung2012}
Srinivasan K, Young WR. 2012.
Zonostrophic instability.
\textit{J. Atmos.\ Sci.} 69:1633--1656

\bibitem[Thomas et~al.(2015)Thomas, Farrell, Ioannou \&
  Gayme]{thomas_etal_2015}
Thomas VL, Farrell BF, Ioannou PJ, Gayme DF. 2015.
A minimal model of self-sustaining turbulence.
\textit{Physics of Fluids} 27(10):105104

\bibitem[Thomas et~al.(2014)Thomas, Lieu, Jovanovi{\'c}, Farrell, Ioannou \&
  Gayme]{thomas_etal_2014}
Thomas VL, Lieu BK, Jovanovi{\'c} MR, Farrell BF, Ioannou PJ, Gayme DF. 2014.
Self-sustaining turbulence in a restricted nonlinear model of plane couette
  flow.
\textit{Physics of Fluids} 26(10):105112

\bibitem[Thomson(1880)]{Thomson:1880bv}
Thomson W. 1880.
{On Gravitational Oscillations of Rotating Water}.
\textit{Proceedings of the Royal Society of Edinburgh} 10:92 -- 100

\bibitem[Tobias(2021)]{Tobias:2021}
Tobias S. 2021.
The turbulent dynamo.
\textit{Journal of fluid mechanics} 912

\bibitem[Tobias et~al.(2011)Tobias, Dagon \& Marston]{tobias2011astrophysical}
Tobias S, Dagon K, Marston J. 2011.
{Astrophysical fluid dynamics via direct statistical simulation}.
\textit{The Astrophysical Journal} 727(2):127

\bibitem[Tobias \& Marston(2013)]{tobias2013direct}
Tobias S, Marston J. 2013.
{Direct statistical simulation of out-of-equilibrium jets}.
\textit{Physical review letters} 110(10):104502

\bibitem[{Tobias} \& {Marston}(2017)]{tm2017}
{Tobias} SM, {Marston} JB. 2017.
{Three-dimensional rotating Couette flow via the generalised quasilinear
  approximation}.
\textit{Journal of Fluid Mechanics} 810:412--428

\bibitem[{Tobias} et~al.(2018){Tobias}, {Oishi} \& {Marston}]{tom_2018}
{Tobias} SM, {Oishi} JS, {Marston} JB. 2018.
{Generalized quasilinear approximation of the interaction of convection and
  mean flows in a thermal annulus}.
\textit{Proceedings of the Royal Society of London Series A} 474(2219):20180422

\bibitem[Touchette(2009)]{Touchette:2009eb}
Touchette H. 2009.
{The large deviation approach to statistical mechanics}.
\textit{Physics Reports} 478(1-3):1 -- 69

\bibitem[Tretiak et~al.(2022)Tretiak, Plumley, Calkins \& Tobias]{Tretiak_2022}
Tretiak K, Plumley M, Calkins M, Tobias S. 2022.
Efficiency gains of a multi-scale integration method applied to a
  scale-separated model for rapidly rotating dynamos.
\textit{Computer Physics Communications} 273:108253

\bibitem[Turton et~al.(2015)Turton, Tuckerman \& Barkley]{turton_etal}
Turton SE, Tuckerman LS, Barkley D. 2015.
Prediction of frequencies in thermosolutal convection from mean flows.
\textit{Phys. Rev. E} 91(4):043009

\bibitem[Vedenov(1963)]{Vedenov:1963jj}
Vedenov AA. 1963.
{Quasi-linear theory of a plasma}.
\textit{Soviet Atomic Energy} 13(1):591 -- 612

\bibitem[Vedenov et~al.(1961)Vedenov, Velikhov \& Sagdeev]{Vedenov:1961us}
Vedenov AA, Velikhov EP, Sagdeev RZ. 1961.
{Quasilinear theory of plasma oscillations}.
\textit{Proceedings of IAEA Conference on Plasma Physics and Controlled Nuclear
  Fusion Research} :465 -- 475

\bibitem[Venaille \& Delplace(2021)]{Venaille2021}
Venaille A, Delplace P. 2021.
{Wave topology brought to the coast}.
\textit{Physical Review Research} 3(4):043002

\bibitem[Venturi(2018)]{Venturi:2018bp}
Venturi D. 2018.
{The numerical approximation of nonlinear functionals and functional
  differential equations}.
\textit{Physics Reports} 732:1 -- 102

\bibitem[Willis \& Kerswell(2007)]{Willis2007}
Willis AP, Kerswell RR. 2007.
{Turbulent dynamics of pipe flow captured in a reduced model: puff
  relaminarization and localized `edge' states}.
\textit{Journal of Fluid Mechanics} 619:213--233

\bibitem[Woillez \& Bouchet(2017)]{Woillez:2017fr}
Woillez E, Bouchet F. 2017.
{Theoretical prediction of Reynolds stresses and velocity profiles for
  barotropic turbulent jets}.
\textit{EPL (Europhysics Letters)} 118(5):54002 -- 7

\bibitem[Yaglom(1994)]{yag_1994}
Yaglom AM. 1994.
A. n. kolmogorov as a fluid mechanician and founder of a school in turbulence
  research.
\textit{Annual Review of Fluid Mechanics} 26(1):1--23

\bibitem[{Yokoi}(2019)]{yokoi19}
{Yokoi} N. 2019.
{Turbulence, transport and reconnection}, In \textit{CISM Courses and Lectures:
  Advanced Topics in MHD}

\bibitem[Zare et~al.(2017)Zare, Jovanovi{\'c} \& Georgiou]{zare_etal_2017}
Zare A, Jovanovi{\'c} MR, Georgiou TT. 2017.
Colour of turbulence.
\textit{Journal of Fluid Mechanics} 812:636--680

\bibitem[Zhang et~al.(2019)Zhang, Lawrence, Marston \&
  Kushner]{zhang2019infinite}
Zhang C, Lawrence A, Marston B, Kushner PJ. 2019.
Infinite u(1) symmetry of the quasi-linear approximation, In \textit{22nd
  Conference on Atmospheric and Oceanic Fluid Dynamics}. AMS

\bibitem[Zhou(2021)]{Zhou2021}
Zhou Y. 2021.
{Turbulence theories and statistical closure approaches}.
\textit{Physics Reports} 935:1--117

\bibitem[Zhu et~al.(2021)Zhu, Li \& Marston]{Zhu2021}
Zhu Z, Li C, Marston JB. 2021.
{Topology of rotating stratified fluids with and without background shear
  flow}.
\textit{arXiv:2112.04691}

\end{thebibliography}

\end{document}